\documentclass[aps,a4paper,superscriptaddress,preprintnumbers,showpacs,
amsmath,amssymb]{revtex4}
\usepackage{graphicx}
\usepackage{dcolumn}
\usepackage{color}
\usepackage{latexsym,amsfonts}
\usepackage{textcomp}
\usepackage{bm}

\usepackage{ulem}

\begin{document}

\title{Effective Low--Energy Potential\\ for Slow Dirac Fermions in
  Einstein--Cartan Gravity with Torsion and Chameleon\footnote{We
    dedicate this paper to 200 Jubilee of Vienna University of
    Technology}}

\author{A. N. Ivanov}\email{ivanov@kph.tuwien.ac.at}
\affiliation{Atominstitut, Technische Universit\"at Wien, Stadionallee
  2, A-1020 Wien, Austria}
\author{M. Wellenzohn}\email{max.wellenzohn@gmail.com}
\affiliation{Atominstitut, Technische Universit\"at Wien, Stadionallee
  2, A-1020 Wien, Austria}
\affiliation{FH Campus Wien, University of Applied Sciences, 
Favoritenstra\ss e 226, 1100 Wien, Austria}

\date{\today}

\begin{abstract}
We derive the most general effective low--energy potential to order
$O(1/m)$ for slow Dirac fermions with mass $m$, coupled to
gravitational, chameleon and torsion fields in the Einstein--Cartan
gravity. The obtained results can be applied to the experimental
analysis of gravitational, chameleon and torsion interactions in
terrestrial laboratories.  We discuss the use of rotating coordinate
systems, caused by rotations of devices, for measurements of the
torsion vector and tensor components, caused by minimal
torsion--fermion couplings (Ivanov and Wellenzohn, Phys. Rev. D {\bf
  92}, 065006 (2015)). Using the most general form of a metric tensor
of curved spacetimes in rotating coordinate systems, proposed by
Obukhov, Silenko, and Teryaev (Phys. Rev. D {\bf 84}, 024025 (2011)),
we extend this metric by the inclusion of the chameleon field and
calculate the set of vierbein fields, in terms of which Dirac fermions
couple to torsion vector and tensor components through minimal
torsion--fermion couplings. For such a set of vierbein fields we
discuss a part of the effective low--energy potential for slow Dirac
fermions, coupled to gravitational, chameleon and torsion fields to
order $O(1)$ in the large fermion mass expansion.
\end{abstract}

\maketitle

\section{Introduction}
\label{sec:introduction}

In terrestrial laboratories \cite{Abele2010}--\cite{Lemmel2015}
gravitational and chameleon interactions are being investigated in
terms of cold and ultracold neutrons through some effective
low--energy potentials \cite{Brax2011}--\cite{Ivanov2015}. The
low--energy torsion--fermion interactions of the pseudoscalar and
axial--vector components of torsion field have been derived and
estimated by L\"ammerzahl \cite{Laemmerzahl1997} and Obukhov, Silenko,
and Teryaev \cite{Obukhov2014}. The most general torsion--fermion
interactions of constant torsion fields have been proposed and
estimated by Kostelecky, Russell, and Tasson
\cite{Kostelecky2008}. The results, obtained by L\"ammerzahl
\cite{Laemmerzahl1997}, Obukhov {\it et al.}  \cite{Obukhov2014} and
Kostelecky {\it et al.}  \cite{Kostelecky2008}, have been discussed in
\cite{Ivanov2015a}. An attempt of a direct measurement of
torsion--fermion interactions with constant torsion fields, proposed
by Kostelecky {\it et al.}  \cite{Kostelecky2008}, has been undertaken
by Lehnert, Snow and Yan \cite{Lehnert2014}.

In this paper we derive the most general effective low--energy
potential to order $1/m$ for slow Dirac fermions with mass $m$,
coupled to gravitational, chameleon and torsion fields in the
Einstein--Cartan gravity.

The chameleon part of such an effective low--energy potential contains
new chameleon--fermion interactions with respect to those calculated
in \cite{Ivanov2014,Ivanov2015a}. These new chameleon--fermion
interactions can be used for more detailed experimental analysis of
the properties of the chameleon field
\cite{Chameleon1,Chameleon2}. The chameleon field, changing its mass
in dependence of a mass density of environment, has been invented to
avoid the problem of violation of the equivalence principle
\cite{Will1993}. In addition the chameleon field can be also
identified with a quintessence (canonical scalar field)
\cite{Zlatev1999,Tsujikawa2013}, which has been postulated for an
explanation of the late--time acceleration of the Universe expansion
\cite{Perlmutter1997,Riess1998,Perlmutter1999}. The laboratory probes
of the chameleon field, coupled to a matter in conformal way
\cite{Chameleon1,Chameleon2}, may also shed light on dark energy
dynamics \cite{Copeland2006,Frieman2008,Brax2015,Pignol2015}.

Torsion is an additional to a metric tensor natural geometrical
quantity characterizing spacetime geometry through spin--matter
interactions \cite{Hehl1995,Hammond2002}.  It allows to probe
rotational degrees of freedom of spacetime in terrestrial
laboratories.  Torsion is described by a third--order tensor ${\cal
  T}_{\sigma\mu\nu}$, antisymmetric with respect to indices $\mu$ and
$\nu$, i.e. ${\cal T}_{\sigma\mu\nu} = - {\cal T}_{\sigma\nu\mu}$
\cite{Hehl1995,Hammond2002}.  It possesses 24 independent components,
which can be decomposed into 4 axial--vector ${\cal B}_{\mu}$, 4
vector ${\cal E}_{\mu}$ and 16 tensor ${\cal M}_{\sigma\mu\nu}$
components \cite{Kostelecky2008} (see also Eq.(\ref{eq:6}) and
Eq.(\ref{eq:7})). The torsion part of the effective low--energy
potential, derived in our paper, is caused by torsion--fermion minimal
couplings for all torsion components only.  An importance of this part
of the effective low--energy potential is related to a possible
solution of the following problem.  As has been shown in
\cite{Ivanov2015a}, only torsion axial--vector ${\cal B}_{\mu}$
components are present in the torsion--fermion minimal couplings in
the curved spacetimes with metrics, providing vanishing time--space
(space--time) components of the vierbein fields. Since these are usual
metrics of spacetimes in terrestrial laboratories, in such spacetimes
torsion vector ${\cal E}_{\mu}$ and tensor ${\cal M}_{\sigma\mu\nu}$
components, coupled to Dirac fermions, can be introduced only through
non--minimal torsion--fermion couplings with phenomenological coupling
constants \cite{Kostelecky2008} (see also \cite{Ivanov2015a}). The
presence of phenomenological coupling constants screens real values of
torsion vector ${\cal E}_{\mu}$ and tensor ${\cal M}_{\sigma\mu\nu}$
components. Thus, a search for possible ways of measurements of
torsion vector ${\cal E}_{\mu}$ and tensor ${\cal M}_{\sigma\mu\nu}$
components through torsion--fermion minimal couplings is of great deal
of importance for understanding of correct values of torsion.

We show that these measurements can be in principle possible in curved
spacetimes, described by metric tensors with non--diagonal
components. These metric tensors define non--vanishing time--space
(space--time) components of the vierbein fields, in terms of which
slow fermions couple to torsion vector ${\cal E}_{\mu}$ and tensor
${\cal M}_{\sigma\mu\nu}$ components through minimal torsion--fermion
couplings.

It is well--known \cite{LL2008} that in rotating coordinate systems
spacetimes are described by non--diagonal metric tensors. In
terrestrial laboratories spacetimes with non--diagonal metric tensors
can be in principle realized by means of rotating devices (neutron
interferometers) \cite{Atwood1984,Mashhoon1988} (see
  also a book by Rauch and Werner \cite{Rauch2015} for a necessary
  information on neutron interferometry. Thus, we propose to measure
torsion vector $\vec{\cal E}$ and tensor ${\cal M}_{\sigma\mu\nu}$
components through minimal torsion--fermion couplings in rotating
coordinate systems, caused by rotating devices. In our analysis of
curved spacetimes in rotating coordinate systems we follow the papers
by Hehl and Ni \cite{Hehl1990} and Obukhov, Silenko, and Teryaev
\cite{Obukhov2009,Obukhov2011}.

The paper is organized as follows. In section \ref{sec:fermion} we
derive the most general Hamilton operator for the Dirac fermions in
the Einstein--Cartan gravity with chameleon and torsion. We adduce the
Schr\"odinger--Pauli equation for slow Dirac fermions, coupled to the
effective low--energy potential, caused by gravitational, chameleon
and torsion fields. In section \ref{sec:conclusion} we i) calculate
the vierbein fields, related to the most general metric tensors of the
curved spacetimes in rotating coordinate systems, proposed by Obukhov,
Silenko, and Teryaev \cite{Obukhov2011}, ii) derive the effective
low--energy torsion--fermion potential to order $O(1)$ in the large
fermion mass expansion and iii) discuss possible measurements of
torsion vector and tensor components in such curved spacetimes. In the
Appendix we give a detailed derivation of the effective low--energy
potential for slow Dirac fermions, coupled to gravitational, chameleon
and torsion fields in the Einstein--Cartan gravity.

\section{Slow Dirac fermions in the Einstein--Cartan gravity with 
torsion and chameleon}
\label{sec:fermion}

In gravitational theories with chameleon field fermions couple to the
chameleon field $\phi(x)$ through the metric $\tilde{g}_{\mu\nu}(x)$
in the Jordan frame related to the metric $g_{\mu\nu}(x)$ in the
Einstein frame by $\tilde{g}_{\mu\nu}(x) = f^2(x)\,g_{\mu\nu}(x)$,
where $f(x) = e^{\,\beta\phi(x)/M_{\rm Pl}}$ is the conformal factor
\cite{Chameleon1,Chameleon2} (see also \cite{Ivanov2015}), $\beta$ is
the chameleon--matter coupling constant and $M_{\rm Pl} = 1/\sqrt{8\pi
  G_N} = 2.435 \times 10^{27}\,{\rm eV}$ is the reduced Planck mass
and $G_N$ is the gravitational coupling \cite{PDG2014}.

For the derivation of the required effective low--energy potential we
start with the analysis of the Dirac equation for fermions with mass
$m$, coupled to the chameleon field in the spacetime with torsion and
the metric $d\tilde{s}^2 = \tilde{g}_{\mu\nu}(x)dx^{\mu}dx^{\nu}$ (the
Jordan--frame metric). The Dirac--fermion action we take in the
following form \cite{Kostelecky2004} (see also \cite{Ivanov2015a})
\begin{eqnarray}\label{eq:1}
\hspace{-0.3in}{\rm S}_{\psi} = \int d^4x\,\sqrt{-
  \tilde{g}}\,\Big(i\,\frac{1}{2}\,\tilde{e}^{\mu}_{\hat{\lambda}}(x)
\bar{\psi}(x)\gamma^{\hat{\lambda}}\!\!
\stackrel{\leftrightarrow}{D}_{\mu}\!\!\psi(x) -
m\bar{\psi}(x)\psi(x)\Big),
\end{eqnarray}
where $\tilde{e}^{\mu}_{\hat{\lambda}}(x)$ and
$\gamma^{\hat{\lambda}}$ are the vierbein fields, mapping the curved
spacetime onto the Minkowski spacetime \cite{Kostelecky2004} (see also
\cite{Ivanov2015,Ivanov2015a}), and the Dirac matrices in the
Minkowski spacetime \cite{Itzykson1980}, respectively. The first term
in the brackets of Eq.(\ref{eq:1}) takes the form $
\tilde{e}^{\mu}_{\hat{\lambda}}(x)\bar{\psi}(x)\gamma^{\hat{\lambda}}
\!\!\! \stackrel{\leftrightarrow}{D}_{\mu}\!\! \psi(x) =
\tilde{e}^{\mu}_{\hat{\lambda}}(x)
(\bar{\psi}(x)\gamma^{\hat{\lambda}}D_{\mu}\psi(x) -
(\bar{\psi}(x)\bar{D}_{\mu})\gamma^{\hat{\lambda}}\psi(x)$
\cite{Kostelecky2004}, where $D_{\mu}\psi(x)$ and
$(\bar{\psi}(x)\bar{D}_{\mu})$ are the covariant derivatives defined
by \cite{Ivanov2015,Ivanov2015a}
\begin{eqnarray}\label{eq:2}
\hspace{-0.3in} D_{\mu}\psi(x) = \partial_{\mu}\psi(x) -
\tilde{\Gamma}_{\mu}(x)\psi(x)\;,\;(\bar{\psi}(x)\bar{D}_{\mu}) =
\partial_{\mu} \bar{\psi}(x) -
\gamma^{\hat{0}}\tilde{\Gamma}^{\dagger}_{\mu}(x)\gamma^{\hat{0}}.
\end{eqnarray}
The spin affine connection $\tilde{\Gamma}_{\mu}(x)$ is given by
\cite{Ivanov2015a}
\begin{eqnarray}\label{eq:3}
\hspace{-0.3in} \tilde{\Gamma}_{\mu}(x) =
\frac{i}{4}\,\tilde{\omega}_{\mu\hat{\alpha}\hat{\beta}}(x)\,
\sigma^{\hat{\alpha}\hat{\beta}},
\end{eqnarray}
where $\sigma^{\hat{\alpha}\hat{\beta}} = (i/2)
(\gamma^{\hat{\alpha}}\gamma^{\hat{\beta}} -
\gamma^{\hat{\beta}}\gamma^{\hat{\alpha}})$ are the Dirac matrices in
the Minkowski spacetime \cite{Itzykson1980} and the spin connection
$\tilde{\omega}_{\mu\hat{\alpha}\hat{\beta}}(x)$ is related to the
vierbein fields $\tilde{e}^{\mu}_{\hat{\lambda}}(x)$ and the affine
connection ${\tilde{\Gamma}^{\alpha}\,}_{\mu\nu}(x) =
\widetilde{\{{^\alpha}_{\mu\nu}\}} + {\tilde{\cal
    K}^{\alpha}\,}_{\mu\nu}(x)$ as follows \cite{Ivanov2015a}
\begin{eqnarray}\label{eq:4}
\tilde{\omega}_{\mu\hat{\alpha}\hat{\beta}}(x) = -
\eta_{\hat{\alpha}\hat{\varphi}}\Big(\partial_{\mu}\tilde{e}^{\hat{\varphi}}_{\nu}(x)
- {\tilde{\Gamma}^{\alpha}\,}_{\mu\nu}(x)
\tilde{e}^{\hat{\varphi}}_{\alpha}(x)\Big)\tilde{e}^{\nu}_{\hat{\beta}}(x),
\end{eqnarray}
where $\widetilde{\{{^\alpha}_{\mu\nu}\}}$ are the Christoffel symbols
\cite{LL2008} 
\begin{eqnarray}\label{eq:5}
\widetilde{\{{^\alpha}_{\mu\nu}\}} = \frac{1}{2}\,\tilde{g}^{\alpha\lambda}\Big(\frac{\partial
  \tilde{g}_{\lambda\mu}}{\partial x^{\nu}} + \frac{\partial
  \tilde{g}_{\lambda\nu}}{\partial x^{\mu}} - \frac{\partial
  \tilde{g}_{\mu\nu}}{\partial x^{\lambda}}\Big)
\end{eqnarray}
and ${\tilde{\cal K}^{\alpha}\,}_{\mu\nu}(x) = - \frac{1}{2}(
{\tilde{\cal T}^{\alpha}\,}_{\mu\nu}(x) - {{\tilde{\cal
      T}_{\mu}\,}^{\alpha}\,}_{\nu}(x) - {{\tilde{\cal
      T}_{\nu}\,}^{\alpha}\,}_{\mu}(x))$ is the contorsion tensor,
expressed in terms of the torsion field ${\tilde{\cal
    T}^{\alpha}\,}_{\mu\nu}(x) =
\tilde{g}^{\alpha\sigma}(x)\,\tilde{\cal T}_{\sigma\mu\nu}(x)$
\cite{Ivanov2015a}. The torsion field $\tilde{\cal
  T}_{\sigma\mu\nu}(x)$ can be represented in the following
irreducible form \cite{Kostelecky2008} (see also \cite{Ivanov2015a})
\begin{eqnarray}\label{eq:6}
\tilde{{\cal T}}_{\sigma\mu\nu}(x) =
\frac{1}{3}\,\Big(\tilde{g}_{\sigma\mu}(x) \tilde{\cal E}_{\nu}(x) -
\tilde{g}_{\sigma\nu}(x) \tilde{\cal E}_{\mu}(x)\Big) +
\frac{1}{3}\,\tilde{\varepsilon}_{\sigma\mu\nu\alpha}(x)\, \tilde{\cal
  B}^{\alpha}(x) + \tilde{\cal M}_{\sigma\mu\nu}(x),
\end{eqnarray}
where the 4--vector $\tilde{\cal E}_{\nu}(x)$ and axial 4--vector
$\tilde{\cal B}^{\alpha}(x)$ fields, possessing 4 independent
components each, are defined by
\begin{eqnarray}\label{eq:7}
\tilde{\cal E}_{\nu}(x) = \tilde{g}^{\sigma\mu}(x)\,\tilde{\cal
  T}_{\sigma\mu\nu}(x)\quad,\quad \tilde{\cal B}^{\alpha}(x) =
\frac{1}{2}\,\tilde{\varepsilon}^{\alpha\sigma\mu\nu}(x)\,\tilde{\cal
  T}_{\sigma\mu\nu}(x).
\end{eqnarray}
Here $\tilde{\varepsilon}_{\sigma\mu\nu\alpha}(x) = \sqrt{-
  \tilde{g}(x)}\,\epsilon_{\sigma\mu\nu\alpha}$ and
$\tilde{\varepsilon}^{\alpha\sigma\mu\nu}(x) =
\epsilon^{\alpha\sigma\mu\nu}/\sqrt{- \tilde{g}(x)}$ are covariant
Levi--Civita tensors in the curved spacetime with the Jordan metric
$\tilde{g}_{\mu\nu}(x)$ and the definition $\epsilon^{0123} = -
\epsilon_{0123} = + 1$ \cite{LL2008}.  For the derivation of the
axial--vector field $\tilde{\cal B}^{\alpha}(x)$ in terms of the
torsion tensor field $\tilde{\cal T}_{\sigma\mu\nu}(x)$ we have used
the relation $\epsilon^{\alpha\sigma\mu\nu}
\epsilon_{\sigma\mu\nu\beta} = - 6\,\delta^{\alpha}_{\beta}$
\cite{Itzykson1980}.

The residual 16 independent components of the torsion field
$\tilde{\cal T}_{\sigma\mu\nu}(x)$ can be attributed to the tensor
field $\tilde{\cal M}_{\sigma\mu\nu}(x)$, which obeys the constraints
$\tilde{g}^{\sigma\mu}(x) \tilde{\cal M}_{\sigma\mu\nu}(x) =
\tilde{\varepsilon}^{\alpha\sigma\mu\nu}(x) \tilde{\cal
  M}_{\sigma\mu\nu}(x) = 0$ \cite{Kostelecky2008}.

The derivation of the Dirac equation in the curved spacetime with the
metric tensor $\tilde{g}_{\mu\nu}(x)$ and torsion we have carried out
in the Appendix of Ref.\cite{Ivanov2015a}. The result is
\begin{eqnarray}\label{eq:8}
\Big( i\,\tilde{e}^{\mu}_{\hat{\lambda}}(x) \gamma^{\hat{\lambda}}
D_{\mu} - \frac{1}{2}\,i\,{\tilde{\cal
    T}^{\alpha}\,\!\!}_{\alpha\mu}(x)
\tilde{e}^{\mu}_{\hat{\lambda}}(x) \gamma^{\hat{\lambda}} -
\frac{1}{2}\,i\, \tilde{\omega}_{\mu\hat{\alpha}\hat{\beta}}(x)
\tilde{e}^{\mu}_{\hat{\lambda}}(x)
\Big(\eta^{\hat{\lambda}\hat{\beta}}\gamma^{\hat{\alpha}} +
\frac{1}{4}\,i\, [\sigma^{\hat{\alpha}\hat{\beta}},
  \gamma^{\hat{\lambda}}]\Big) - m\Big)\,\psi(x) = 0,
\end{eqnarray}
where $[\sigma^{\hat{\alpha}\hat{\beta}}, \gamma^{\hat{\lambda}}] =
\sigma^{\hat{\alpha}\hat{\beta}} \gamma^{\hat{\lambda}} -
\gamma^{\hat{\lambda}} \sigma^{\hat{\alpha}\hat{\beta}}=
2\,i\,(\eta^{\hat{\beta}\hat{\lambda}}\,\gamma^{\hat{\alpha}} -
\eta^{\hat{\lambda}\hat{\alpha}}\,\gamma^{\hat{\beta}}) $
\cite{Ivanov2015a}. The Dirac equation Eq.(\ref{eq:8}) agrees well
with that derived by Kostelecky (see Eq.(18) of
Ref.\cite{Kostelecky2004}).  The vierbein fields
$\tilde{e}^{\hat{\alpha}}_{\mu}(x)$ and
$\tilde{e}^{\mu}_{\hat{\alpha}}(x)$ in the Jordan frame are related to
the vierbein fields $e^{\hat{\alpha}}_{\mu}(x)$ and
$e^{\mu}_{\hat{\alpha}}(x)$ in the Einstein frame by \cite{Ivanov2015}
\begin{eqnarray}\label{eq:9}
\tilde{e}^{\hat{\alpha}}_{\mu}(x) = f(x)\,e^{\hat{\alpha}}_{\mu}(x)\;,\;
    \tilde{e}^{\mu}_{\hat{\alpha}}(x) = e^{\mu}_{\hat{\alpha}}(x)/f(x).
\end{eqnarray}
The Dirac equation Eq.(\ref{eq:8}) in its standard form reads
\begin{eqnarray}\label{eq:10}
i\,\frac{\partial \psi(t,\vec{r}\,)}{\partial t} = {\rm
  H}\,\psi(t,\vec{r}\,),
\end{eqnarray}
where $\psi(t,\vec{r}\,)$ is the Dirac wave function of the fermion
with mass $m$. The Hamilton operator ${\rm H}$ in its
non--perturbative form is given by \cite{Ivanov2015a}
\begin{eqnarray}\label{eq:11}
{\rm H} &=& \tilde{E}^{\hat{0}}_0(x)\,\gamma^{\hat{0}}\,m -
\tilde{E}^{\hat{0}}_0(x)\,\tilde{e}^0_{\hat{j}}(x)\,i\,\gamma^{\hat{0}}\,
\gamma^{\hat{j}}\,\frac{\partial}{\partial t} -
\tilde{E}^{\hat{0}}_0(x)\,\tilde{e}^j_{\hat{\lambda}}(x)\,i\,\gamma^{\hat{0}}\,
\gamma^{\hat{\lambda}}\,\frac{\partial}{\partial
  x^j}\nonumber\\ &+&\frac{1}{2}\,i\,\tilde{E}^{\hat{0}}_0(x)\,
\gamma^{\hat{0}}\, \Big({\tilde{\cal
    T}^{\alpha}\,}_{\alpha\mu}(x)\,\tilde{e}^{\mu}_{\hat{\lambda}}(x)\,
\gamma^{\hat{\lambda}} +
\tilde{\omega}_{\mu\hat{\alpha}\hat{\beta}}(x)\,\tilde{e}^{\mu}_{\hat{\lambda}}(x)\,
\eta^{\hat{\lambda}\hat{\beta}}\,\gamma^{\hat{\alpha}}\Big)\nonumber\\ &+&
\frac{1}{4}\,\tilde{E}^{\hat{0}}_0(x)\,\tilde{\omega}_{\mu\hat{\alpha}\hat{\beta}}(x)\,
\tilde{e}^{\mu}_{\hat{\lambda}}(x)\,\epsilon^{\hat{\lambda}\hat{\alpha}\hat{\beta}\hat{\rho}}\,
\gamma^{\hat{0}}\, \gamma_{\hat{\rho}}\,\gamma^5,
\end{eqnarray}
where the vierbein field $\tilde{E}^{\hat{0}}_0(x)$ is defined by
$\tilde{E}^{\hat{0}}_0(x) = \tilde{e}^{\hat{0}}_0(x)/(1 -
\tilde{e}^{\hat{0}}_j(x)\tilde{e}^j_{\hat{0}}(x)) =
\tilde{e}^{\hat{0}}_0(x)/(1 -
\tilde{e}^0_{\hat{j}}(x)\tilde{e}^{\hat{j}}_0(x)) =
1/\tilde{e}^0_{\hat{0}}(x)$. The definition for the vierbein field
$\tilde{E}^{\hat{0}}_0(x)$ follows from the relations
$\tilde{e}^{\mu}_{\hat{\alpha}}(x)\,\tilde{e}^{\hat{\beta}}_{\mu}(x) =
\delta^{\hat{\beta}}_{\hat{\alpha}}$ and
$\tilde{e}^{\mu}_{\hat{\alpha}}(x)\,\tilde{e}^{\hat{\alpha}}_{\nu}(x)
= \delta^{\mu}_{\nu}$. In addition we have used that
$\{\sigma^{\hat{\alpha}\hat{\beta}}, \gamma^{\hat{\lambda}}\} =
\sigma^{\hat{\alpha}\hat{\beta}} \gamma^{\hat{\lambda}} +
\gamma^{\hat{\lambda}} \sigma^{\hat{\alpha}\hat{\beta}}= - 2\,
\epsilon^{\hat{\lambda}\hat{\alpha}\hat{\beta}\hat{\rho}}\gamma_{\hat{\rho}}
\gamma^5$ \cite{Ivanov2015a} and $\gamma^5 =
i\,\gamma^{\hat{0}}\gamma^{\hat{1}}\gamma^{\hat{2}}\gamma^{\hat{3}}$
\cite{Itzykson1980}. For the derivation of the effective low--energy
potential it is convenient to transcribe the Hamilton operator
Eq.(\ref{eq:11}) into the form
\begin{eqnarray}\label{eq:12}
{\rm H} &=& \tilde{E}^{\hat{0}}_0(x)\,\gamma^{\hat{0}} m+
\frac{1}{2}\,i\,\tilde{E}^{\hat{0}}_0(x)\, \Big({\tilde{\cal
    T}^{\alpha}\,\!\!}_{\alpha\mu}(x)\tilde{e}^{\mu}_{\hat{0}}(x) +
\tilde{\omega}_{\mu\hat{0}\hat{\beta}}(x)\,\tilde{e}^{\mu}_{\hat{\lambda}}(x)\,
\eta^{\hat{\lambda}\hat{\beta}}\Big) \nonumber\\ &+&
\frac{1}{4}\,\tilde{E}^{\hat{0}}_0(x)\,\Big(\tilde{\omega}_{\mu
  \hat{j}\hat{k}}(x)\,\tilde{e}^{\mu}_{\hat{0}}(x) +
\tilde{\omega}_{\mu [\hat{0}\hat{j}]}(x)\,
\tilde{e}^{\mu}_{\hat{k}}(x)\Big)\,
\epsilon^{\hat{j}\hat{k}\hat{\ell}}\Sigma_{\hat{\ell}} -
\tilde{E}^{\hat{0}}_0(x)\,\tilde{e}^j_{\hat{0}}(x)\,i\,
\frac{\partial}{\partial x^j} \nonumber\\ && -
\tilde{E}^{\hat{0}}_0(x)\,\tilde{e}^j_{\hat{j}}(x)
\,i\,\gamma^{\hat{0}}\gamma^{\hat{j}}\frac{\partial }{\partial x^j} -
\tilde{E}^{\hat{0}}_0(x)\,\tilde{e}^0_{\hat{j}}(x)\,i\,\gamma^{\hat{0}}\,
\gamma^{\hat{j}}\,\frac{\partial}{\partial t}\nonumber\\&+&
\frac{1}{2}\,\tilde{E}^{\hat{0}}_0(x)\,\Big({\tilde{\cal
    T}^{\alpha}\,\!\!}_{\alpha\mu}(x) \tilde{e}^{\mu}_{\hat{j}}(x) +
\tilde{\omega}_{\mu\hat{j}\hat{\beta}}(x)
\tilde{e}^{\mu}_{\hat{\lambda}}(x)\,\eta^{\hat{\lambda}\hat{\beta}}\Big)\,
i\,\gamma^{\hat{0}}\gamma^{\hat{j}} -
\frac{1}{4}\,\tilde{E}^{\hat{0}}_0(x)\,\tilde{\omega}_{\mu
  \hat{j}\hat{k}}(x)\,
\tilde{e}^{\mu}_{\hat{\ell}}(x)\,\epsilon^{\hat{j}\hat{k}\hat{\ell}}\,
\gamma^5,
\end{eqnarray}
where we have denoted $\tilde{\omega}_{\mu [\hat{0}\hat{j}]}(x) =
\tilde{\omega}_{\mu \hat{0}\hat{j}}(x) - \tilde{\omega}_{\mu
  \hat{j}\hat{0}}(x)$ and used
$\epsilon^{\hat{\ell}\hat{j}\hat{k}\hat{0}} = -
\epsilon^{\hat{j}\hat{k}\hat{\ell}}$ \cite{Itzykson1980}. It is
well--known \cite{Obukhov2014} that the Hamilton operator
Eq.(\ref{eq:12}) is not hermitian. This is because of the factor
$\sqrt{-\tilde{g}}$ in the definition of the 4--dimensional covariant
volume element $d^4x\,\sqrt{-\tilde{g}}$ in the curved spacetime. In
order to deal with the hermitian Hamilton operator we have to make the
following transformation of the fermion wave function and the Hamilton
operator \cite{Obukhov2014}
\begin{eqnarray}\label{eq:13}
\psi(x) &=& \Big(\sqrt{-
  \tilde{g}(x)}\,\tilde{e}^0_{\hat{0}}(x)\Big)^{-
  1/2}\,\psi'(x),\nonumber\\ {\rm H}' &=& \Big(\sqrt{-
  \tilde{g}(x)}\,\tilde{e}^0_{\hat{0}}(x)\Big)^{+1/2}\,{\rm
  H}\Big(\sqrt{- \tilde{g}(x)}\,\tilde{e}^0_{\hat{0}}(x)\Big)^{- 1/2}
- i\,\Big(\sqrt{-
  \tilde{g}(x)}\,\tilde{e}^0_{\hat{0}}(x)\Big)^{+1/2}\,
\frac{\partial}{\partial t}\Big(\sqrt{-
  \tilde{g}(x)}\,\tilde{e}^0_{\hat{0}}(x)\Big)^{- 1/2}.
\end{eqnarray}
We would like to note that the last term in the
  Hamilton operator ${\rm H}'$ acts on the fermion wave function as a
  multiplication operator and does not differentiate it with respect
  to time. This means that we do not need to add in the definition of
  the Hamilton operator ${\rm H}'$ a time derivative operator
  $i\,\partial/\partial t$ \cite{Neznamov2009,Silenko2013,
    Silenko2015}. The Dirac equation for the wave function
$\psi'(t,\vec{r}\,)$ retains its standard form
\begin{eqnarray}\label{eq:14}
i\,\frac{\partial \psi'(t,\vec{r}\,)}{\partial t} = {\rm
  H}'\,\psi'(t,\vec{r}\,),
\end{eqnarray}
where the hermitian Hamilton operator ${\rm H}'$ is given by
\begin{eqnarray}\label{eq:15}
&&{\rm H}' = \tilde{E}^{\hat{0}}_0(x)\,\gamma^{\hat{0}} m +
  \frac{1}{2}\,i\, \tilde{E}^{\hat{0}}_0(x)\,\Big({\tilde{\cal
      T}^{\alpha}\,\!\!}_{\alpha\mu}(x)\tilde{e}^{\mu}_{\hat{0}}(x) +
  \tilde{\omega}_{\mu\hat{0}\hat{\beta}}(x)\,\tilde{e}^{\mu}_{\hat{\lambda}}(x)\,
  \eta^{\hat{\lambda}\hat{\beta}}\Big) +
  \frac{1}{2}\,i\,\tilde{E}^{\hat{0}}_0(x)\,\frac{1}{\sqrt{-
      \tilde{g}(x)}}\frac{\partial}{\partial t}\Big(\sqrt{-
    \tilde{g}(x)}\,\tilde{e}^0_{\hat{0}}(x)\Big)\nonumber\\ && +
  \frac{1}{4}\,\tilde{E}^{\hat{0}}_0(x)\,\Big(\tilde{\omega}_{\mu
    \hat{j}\hat{k}}(x)\,\tilde{e}^{\mu}_{\hat{0}}(x) +
  \tilde{\omega}_{\mu [\hat{0}\hat{j}]}(x)\,
  \tilde{e}^{\mu}_{\hat{k}}(x)\Big)\,
  \varepsilon^{\hat{j}\hat{k}\hat{\ell}}\Sigma_{\hat{\ell}} +
  \frac{1}{2}\,i\,(\tilde{E}^{\hat{0}}_0(x))^2\,\tilde{e}^j_{\hat{0}}(x)
  \,\frac{1}{\sqrt{- \tilde{g}(x)}}\frac{\partial }{\partial
    x^j}\Big(\sqrt{-
    \tilde{g}(x)}\,\tilde{e}^0_{\hat{0}}(x)\Big)\nonumber\\ && -
  \tilde{E}^{\hat{0}}_0(x)\,\tilde{e}^j_{\hat{0}}(x)\,i\,
  \frac{\partial}{\partial x^j} -
  \tilde{E}^{\hat{0}}_0(x)\,\tilde{e}^j_{\hat{j}}(x)
  \,i\,\gamma^{\hat{0}}\gamma^{\hat{j}}\frac{\partial }{\partial x^j}
  -
  \tilde{E}^{\hat{0}}_0(x)\,\tilde{e}^0_{\hat{j}}(x)\,i\,\gamma^{\hat{0}} 
\gamma^{\hat{j}}\,\frac{\partial}{\partial
    t}\nonumber\\ && +
  \frac{1}{2}\,(\tilde{E}^{\hat{0}}_0(x))^2\,\tilde{e}^j_{\hat{j}}(x)
  \,i\,\gamma^{\hat{0}}\gamma^{\hat{j}}\frac{1}{\sqrt{-
      \tilde{g}(x)}}\frac{\partial }{\partial x^j}\Big(\sqrt{-
    \tilde{g}(x)}\,\tilde{e}^0_{\hat{0}}(x)\Big) +
  \frac{1}{2}\,(\tilde{E}^{\hat{0}}_0(x))^2\,\tilde{e}^0_{\hat{j}}(x)
  \,i\,\gamma^{\hat{0}}\gamma^{\hat{j}}\,\frac{1}{\sqrt{-
      \tilde{g}(x)}}\,\frac{\partial }{\partial t}\Big(\sqrt{-
    \tilde{g}(x)}\,\tilde{e}^0_{\hat{0}}(x)\Big)\nonumber\\&&+
  \frac{1}{2}\,\tilde{E}^{\hat{0}}_0(x)\,\Big({\tilde{\cal
      T}^{\alpha}\,\!\!}_{\alpha\mu}(x) \tilde{e}^{\mu}_{\hat{j}}(x) +
  \tilde{\omega}_{\mu\hat{j}\hat{\beta}}(x)
  \tilde{e}^{\mu}_{\hat{\lambda}}(x)\,\eta^{\hat{\lambda}\hat{\beta}}\Big)\,
  i\,\gamma^{\hat{0}}\gamma^{\hat{j}} -
  \frac{1}{4}\,\tilde{\omega}_{\mu
    \hat{j}\hat{k}}(x)\,\tilde{E}^{\hat{0}}_0(x)
  \tilde{e}^{\mu}_{\hat{\ell}}(x)\,\varepsilon^{\hat{j}\hat{k}\hat{\ell}}\,
  \gamma^{\hat{5}}.
\end{eqnarray}
Using the relation (see Eq.(A-9) of Ref.\cite{Ivanov2015a})
\begin{eqnarray}\label{eq:16}
\frac{1}{2}\,i\,\frac{1}{\sqrt{- \tilde{g}}}\,\frac{\partial
}{\partial t}\Big(\sqrt{- \tilde{g}}\,\tilde{e}^0_{\hat{0}}(x)\Big) =
- \frac{1}{2}\,i\,{{\cal \tilde{T}}^{\alpha}\,}_{\alpha\mu}(x)
\tilde{e}^{\mu}_{\hat{0}}(x) - \frac{1}{2}\,i\,
\tilde{\omega}_{\mu\hat{0}\hat{\beta}}(x)\tilde{e}^{\mu}_{\hat{\lambda}}(x)\,
\eta^{\hat{\lambda}\hat{\beta}} - \frac{1}{2}\,i\,\frac{1}{\sqrt{-
    \tilde{g}}}\,\frac{\partial }{\partial x^j}\Big(\sqrt{-
  \tilde{g}}\,\tilde{e}^j_{\hat{0}}(x)\Big)
\end{eqnarray}
we transcribe the Hamilton operator Eq.(\ref{eq:15}) into the form
\begin{eqnarray}\label{eq:17}
\hspace{-0.3in}&&{\rm H}' = \tilde{E}^{\hat{0}}_0(x)\,\gamma^{\hat{0}}
m - \frac{1}{2}\,i\,\tilde{E}^{\hat{0}}_0(x)\,\frac{1}{\sqrt{-
    \tilde{g}}}\,\frac{\partial }{\partial x^j}(\sqrt{-
  \tilde{g}}\,\tilde{e}^j_{\hat{0}}(x)) +
\frac{1}{2}\,i\,(\tilde{E}^{\hat{0}}_0(x))^2\,\tilde{e}^j_{\hat{0}}(x)
\,\frac{1}{\sqrt{- \tilde{g}(x)}}\frac{\partial }{\partial
  x^j}\Big(\sqrt{-
  \tilde{g}(x)}\,\tilde{e}^0_{\hat{0}}(x)\Big)\nonumber\\ \hspace{-0.3in}&&
+ \frac{1}{4}\,\tilde{E}^{\hat{0}}_0(x)\,\Big(\tilde{\omega}_{\mu
  \hat{j}\hat{k}}(x)\,\tilde{e}^{\mu}_{\hat{0}}(x) +
\tilde{\omega}_{\mu [\hat{0}\hat{j}]}(x)\,
\tilde{e}^{\mu}_{\hat{k}}(x)\Big)\,
\epsilon^{\hat{j}\hat{k}\hat{\ell}}\Sigma_{\hat{\ell}} -
\tilde{E}^{\hat{0}}_0(x)\,\tilde{e}^j_{\hat{0}}(x)\,i\,
\frac{\partial}{\partial x^j}\nonumber\\ 
\hspace{-0.3in}&& - \tilde{E}^{\hat{0}}_0(x)\,\tilde{e}^j_{\hat{j}}(x)
\,i\,\gamma^{\hat{0}}\gamma^{\hat{j}}\frac{\partial }{\partial x^j} -
\tilde{E}^{\hat{0}}_0(x)\,\tilde{e}^0_{\hat{j}}(x)\,i\,\gamma^{\hat{0}}
\gamma^{\hat{j}}\,\frac{\partial}{\partial t} +
\frac{1}{2}\,(\tilde{E}^{\hat{0}}_0(x))^2\,\tilde{e}^j_{\hat{j}}(x)
\,i\,\gamma^{\hat{0}}\gamma^{\hat{j}}\frac{1}{\sqrt{-
    \tilde{g}(x)}}\frac{\partial }{\partial x^j}\Big(\sqrt{-
  \tilde{g}(x)}\,\tilde{e}^0_{\hat{0}}(x)\Big)\nonumber\\ \hspace{-0.3in}&&
- \frac{1}{2}\,(\tilde{E}^{\hat{0}}_0(x))^2\,\tilde{e}^0_{\hat{j}}(x)
\,i\,\gamma^{\hat{0}}\gamma^{\hat{j}}\,\frac{1}{\sqrt{-
    \tilde{g}(x)}}\,\frac{\partial }{\partial x^j}\Big(\sqrt{-
  \tilde{g}(x)}\,\tilde{e}^j_{\hat{0}}(x)\Big) +
\frac{1}{2}\,\tilde{E}^{\hat{0}}_0(x)\,\Big({\tilde{\cal
    T}^{\alpha}\,\!\!}_{\alpha\mu}(x) \tilde{e}^{\mu}_{\hat{j}}(x) +
\tilde{\omega}_{\mu\hat{j}\hat{\beta}}(x)
\tilde{e}^{\mu}_{\hat{\lambda}}(x)\,\eta^{\hat{\lambda}\hat{\beta}}\Big)\,
i\,\gamma^{\hat{0}}\gamma^{\hat{j}}\nonumber\\
\hspace{-0.3in}&& -
\frac{1}{2}\,(\tilde{E}^{\hat{0}}_0(x))^2\,\tilde{e}^0_{\hat{j}}(x)
\,\Big({{\cal \tilde{T}}^{\alpha}\,}_{\alpha\mu}(x)
\tilde{e}^{\mu}_{\hat{0}}(x) +
\tilde{\omega}_{\mu\hat{0}\hat{\beta}}(x)\tilde{e}^{\mu}_{\hat{\lambda}}(x)\,
\eta^{\hat{\lambda}\hat{\beta}}\Big)\,i\,\gamma^{\hat{0}}\gamma^{\hat{j}}\,
- \frac{1}{4}\,\tilde{\omega}_{\mu
  \hat{j}\hat{k}}(x)\,\tilde{E}^{\hat{0}}_0(x)\,
\tilde{e}^{\mu}_{\hat{\ell}}(x)\,\epsilon^{\hat{j}\hat{k}\hat{\ell}}\,\gamma^{\hat{5}}.
\end{eqnarray}
According to the Foldy--Wouthuysen classification \cite{Foldy1950},
the operators in the first two lines of Eq.(\ref{eq:17}) are {\it
  even}, whereas all other operators are {\it odd}. For a derivation
of a low--energy effective Hamilton operator of slow fermions all {\it
  odd} operators should be removed by some unitary transformations
\cite{Foldy1950}. Skipping standard intermediate Foldy--Wouthuysen
calculations, which are given in the Appendix, we arrive at the
Schr\"odinger--Pauli equation
\begin{eqnarray}\label{eq:18}
i\,\frac{\partial \Psi(t,\vec{r}\,)}{\partial t} = \Big( -
\frac{1}{2m}\,\Delta + \Phi_{\rm
  eff}(t,\vec{r},\vec{\sigma}\,)\Big)\,\Psi(t,\vec{r}\,),
\end{eqnarray}
where $\Psi(t,\vec{r}\,)$ is the large component of the Dirac wave
function of slow fermions with mass $m$ and $\Phi_{\rm
  eff}(t,\vec{r},\vec{\sigma}\,)$ is the effective low--energy
potential for slow fermions, coupled to gravitational, chameleon and
torsion fields, and $\vec{\sigma}$ are the $2\times 2$ Pauli matrices
\cite{Itzykson1980}.  The exact expression of the potential $\Phi_{\rm
  eff}(t,\vec{r},\vec{\sigma}\,)$, calculated to order $O(1/m)$, is
given by Eq.(\ref{eq:A.15}) of the Appendix.

\section{Conclusive discussion}
\label{sec:conclusion}

We have analysed the low--energy approximation of the Dirac equation
for fermions with mass $m$ in the Einstein--Cartan gravity with
torsion and chameleon. Using the Foldy--Wouthuysen transformations we
have derived the most general low--energy potential to order $1/m$ for
slow Dirac fermions, coupled to gravitational, torsion and chameleon
fields. The aim of the derivation of such an effective low--energy
potential is addressed to the investigation of spacetimes in which
torsion vector $\vec{\cal E}$ and tensor ${\cal M}_{jk\ell}$
components, coupled minimally to slow Dirac fermions, can be in
principle observable. As has been shown in \cite{Ivanov2015a} for
metric tensors, yielding vanishing non--diagonal time--space
(space--time) components of the vierbein fields, in the perturbative
regime for gravitational, torsion and chameleon fields only torsion
axial--vector components survive in the low--energy approximation of
the minimal torsion--fermion couplings. 

In the Appendix for the derivation of the effective low--energy
potential $\Phi_{\rm eff}(t, \vec{r}, \vec{\sigma}\,)$ (see
Eq.(\ref{eq:A.15})) we have introduced the operators $A$, $B$,
$C^{\hat{\ell}}$, $D^j_{\hat{j}}$, $F_{\hat{j}}$, $G_{\hat{j}}$, $K$
and $L^j$, which are defined in Eq.(\ref{eq:A.2}).  In the
approximation \cite{Ivanov2015a} applied to our approach these
operators behave as follows
\begin{eqnarray}\label{eq:19}
A &=& 1 + O(g, \phi)\;,\; B = 0\;,\;C^{\hat{\ell}}
= \frac{1}{4}\,{\cal B}^{\hat{\ell}} +
O(g,\phi)\;,\; D^j_{\hat{j}} = \delta^j_{\hat{j}} +
    O(g,\phi),\nonumber\\ F_{\hat{j}} &=& 0\;,\,G_{\hat{j}} =
        O(g, \phi)\;,\; K = - \frac{1}{4}\,{\cal K} + O(g,
        \phi)\;,\; L^j = 0,
\end{eqnarray}
where ${\cal B}^{\hat{\ell}} = \frac{1}{2}\,\epsilon^{\hat{\ell}
  \hat{j}\hat{k} }({\cal T}_{\hat{j} \hat{k} \hat{0}} + {\cal
  T}_{\hat{k} \hat{0} \hat{j}} + {\cal T}_{\hat{0} \hat{j} \hat{k}})$
and ${\cal K} = \frac{1}{2}\,\epsilon^{\hat{j}\hat{k}\hat{\ell}}{\cal
  T}_{\hat{\ell}\hat{j}\hat{k}}$ are the torsion axial--vector and
pseudoscalar components \cite{Ivanov2015a}, and $O(g, \phi)$ are the
linear order contributions of gravitational and chameleon fields. The
non--trivial linear order contributions of the torsion vector and
tensor components can appear only in spacetimes with non--diagonal
metric tensors, yielding non--vanishing non--diagonal time--space
(space--time) components of the vierbein fields. It is well--known
that in the rotating coordinate system spacetime is described by a
non--diagonal metric tensor \cite{LL2008} with the non--vanishing
time--space (space--time) components $g_{0j}(x)$ proportional to the
angular velocity (see also \cite{Hehl1990,Obukhov2009,Obukhov2011}).

The phase--shift induced by a rotational motion of an optical
interferometer was first proposed by Sagnac \cite{Sagnac1913} and
observed by Michelson, Gale, and Pearson \cite{Michelson1925}. In
spite of the fact that the inertial properties of photons and neutrons
are different, the analogous effect for the phase--shift of slow
neutrons was predicted by Page \cite{Page1975} and measured by Werner
{\it et al.}  \cite{Werner1979}, Atwood {\it et al.}
\cite{Atwood1984} and Mashhoon \cite{Mashhoon1988}. For the
measurement of such a phase--shift Atwood {\it et al.}
\cite{Atwood1984} and Mashhoon \cite{Mashhoon1988} used the rotating
two--crystal neutron interferometer and the neutron interferometer in
the rotating reference frame, respectively. According to an
equivalence between a rotating coordinate system and a gravitational
field or a curved spacetime with a corresponding metric tensor
\cite{LL2008}, the experimental setup of the experiments by Atwood
     {\it et al.}  \cite{Atwood1984} and Mashhoon \cite{Mashhoon1988}
     should determine metric tensors of curved spacetimes, created by
     rotating neutron interferometers in the gravitational field of
     the Earth.

Following such an equivalence, for the experimental analysis of
fermion--torsion interactions, described by the effective low--energy
potential $\Phi_{\rm eff}(t, \vec{r}, \vec{\sigma}\,)$ given by
Eq.(\ref{eq:A.15}), we have to determine the metric tensor of the
curved spacetime and calculate the vierbein fields, caused by the
experimental setup of possible experiments. The line element in
spacetimes, created by rotating devices in an arbitrary gravitational
field, we take in the most general form, proposed by Obukhov, Silenko,
and Teryaev \cite{Obukhov2011}:
\begin{eqnarray}\label{eq:20}
d\tilde{s}^2 = \tilde{V}^2(x)\,dt^2 + \eta_{\hat{j}\hat{\ell}}
\tilde{W}^{\hat{j}}_j(x) \tilde{W}^{\hat{\ell}}_{\ell}(x) \Big(dx^j -
K^j(x)\, dt\Big)\Big(dx^{\ell} - K^{\ell}(x)\, dt\Big),
\end{eqnarray}
where the functions $ \tilde{V}^2(x)$ and $\tilde{W}^{\hat{j}}_j(x)$
are defined by an arbitrary gravitational field. In comparison with
Obukhov {\it et al.} \cite{Obukhov2011} they are modified by the
chameleon field. In turn, the functions $K^j(x)$ are caused by
rotations.  The components of the metric tensor
$\tilde{g}_{\mu\nu}(x)$ are equal to
\begin{eqnarray}\label{eq:21}
\tilde{g}_{00}(x) &=& \tilde{V}^2(x) + \eta_{\hat{j}\hat{\ell}}
\tilde{W}^{\hat{j}}_j(x) \tilde{W}^{\hat{\ell}}_{\ell}(x)
K^j(x)K^{\ell}(x)\;,\;\tilde{g}_{0j}(x) = - \eta_{\hat{j}\hat{\ell}}
\tilde{W}^{\hat{j}}_j(x) \tilde{W}^{\hat{\ell}}_{\ell}(x)
K^{\ell}(x),\nonumber\\ \tilde{g}_{j\ell}(x) &=& \eta_{\hat{j}\hat{\ell}}
\tilde{W}^{\hat{j}}_j(x) \tilde{W}^{\hat{\ell}}_{\ell}(x).
\end{eqnarray}
The vierbein fields $\tilde{e}^{\hat{\alpha}}_{\mu}(x)$ are defined by
the relation \cite{Ivanov2015,Ivanov2015a}
\begin{eqnarray}\label{eq:22}
\tilde{g}_{\mu\nu}(x) = \eta_{\hat{\alpha}\hat{\beta}}
\tilde{e}^{\hat{\alpha}}_{\mu}(x)\tilde{e}^{\hat{\beta}}_{\mu}(x).
\end{eqnarray}
Solving Eq.(\ref{eq:22}) for $\tilde{g}_{00}(x)$ and $\tilde{g}_{j\ell}(x)$ we get 
\begin{eqnarray}\label{eq:23}
\tilde{e}^{\hat{0}}_0(x) &=& \sqrt{\tilde{V}^2(x) + (1 -
  \xi)\,\eta_{\hat{j}\hat{\ell}} \tilde{W}^{\hat{j}}_j(x)
  \tilde{W}^{\hat{\ell}}_{\ell}(x) K^j(x) K^{\ell}(x)}\;,\;
\tilde{e}^{\hat{j}}_0(x) =  - \sqrt{\xi}\,\tilde{W}^{\hat{j}}_j(x)K^j(x),\nonumber\\
\tilde{e}^{\hat{0}}_j(x) &=& 0\;,\;
\tilde{e}^{\hat{j}}_j(x) = \tilde{W}^{\hat{j}}_j(x),
\end{eqnarray}
where $\xi$ is a parameter, which can be fixed from Eq.(\ref{eq:21})
for $\tilde{g}_{0j}(x)$. Indeed, using the vierbein fields
Eq.(\ref{eq:23}) we obtain
\begin{eqnarray}\label{eq:24}
\tilde{g}_{0j}(x) &=& - \sqrt{\xi}\,\eta_{\hat{j}\hat{\ell}}
\tilde{W}^{\hat{j}}_j(x) \tilde{W}^{\hat{\ell}}_{\ell}(x) K^{\ell}(x).
\end{eqnarray}
From the comparison of Eq.(\ref{eq:24}) with Eq.(\ref{eq:21}) we
obtain $\xi = 1$. This gives the vierbein fields, given by
Eq.(\ref{eq:23}), equal to
\begin{eqnarray}\label{eq:25}
\tilde{e}^{\hat{0}}_0(x) = \tilde{V}(x)\;,\;\tilde{e}^{\hat{j}}_0(x) =
- \tilde{W}^{\hat{j}}_j(x)K^j(x)\;,\;\tilde{e}^{\hat{0}}_j(x) =
0\;,\;\tilde{e}^{\hat{j}}_j(x) = \tilde{W}^{\hat{j}}_j(x).
\end{eqnarray}
For the calculation of the vierbein fields
$\tilde{e}^{\mu}_{\hat{\alpha}}(x)$ we use the relations
\cite{Ivanov2015,Ivanov2015a}
\begin{eqnarray}\label{eq:26}
\delta^{\hat{\alpha}}_{\hat{\beta}} =
\tilde{e}^{\hat{\alpha}}_{\mu}(x)\,
\tilde{e}^{\mu}_{\hat{\beta}}(x)\;,\;\delta^{\mu}_{\nu} =
\tilde{e}^{\mu}_{\hat{\alpha}}(x)\,\tilde{e}^{\hat{\alpha}}_{\nu}(x).
\end{eqnarray}
Skipping intermediate calculations we obtain
\begin{eqnarray}\label{eq:27}
\tilde{e}^0_{\hat{0}}(x) = \frac{1}{\tilde{V}(x)}\;,\;
\tilde{e}^0_{\hat{j}}(x) = 0\;,\;
\tilde{e}^j_{\hat{0}}(x) = \frac{K^j(x)}{\tilde{V}(x)}\;,\;
\tilde{e}^j_{\hat{j}}(x) = W^j_{\hat{j}}(x).
\end{eqnarray}
The vierbein fields in Eq.(\ref{eq:25}) and Eq.(\ref{eq:27}) have been
calculated at the assumption that the functions $W^{\hat{j}}_j(x)$ and
$W^j_{\hat{j}}(x)$ obey the orthogonality relations
\begin{eqnarray}\label{eq:28}
W^{\hat{j}}_j(x)W^j_{\hat{\ell}} = \delta^{\hat{j}}_{\hat{\ell}}\;,\;
W^j_{\hat{j}}(x)W^{\hat{j}}_{\ell} = \delta^j_{\ell},
\end{eqnarray}
which are fulfilled for the Schwarzschild metric in the weak
gravitational field of the Earth approximation \cite{Ivanov2015a}. For
the verification of the correctness of the obtained vierbein fields we
construct the metric tensor $\tilde{g}^{\mu\nu}(x)$. In terms of the
vierbein fields $\tilde{e}^{\mu}_{\hat{\alpha}}(x)$ it is determined
by
\begin{eqnarray}\label{eq:29}
\tilde{g}^{\mu\nu}(x) =
\eta^{\hat{\alpha}\hat{\beta}}\tilde{e}^{\mu}_{\hat{\alpha}}(x)
\tilde{e}^{\nu}_{\hat{\beta}}(x).
\end{eqnarray}
Using the vierbein fields Eq.(\ref{eq:27}) for the components of the
metric tensor $\tilde{g}^{\mu\nu}(x)$ we obtain the following
expressions
\begin{eqnarray}\label{eq:30}
\tilde{g}^{00}(x) &=&
\frac{1}{\tilde{V}^2(x)}\;,\;\tilde{g}^{0j}(x) =
\frac{K^j(x)}{\tilde{V}(x)},\nonumber\\ \tilde{g}^{j\ell}(x) &=&
\frac{K^j(x)K^{\ell}(x)}{\tilde{V}^2(x)} + \eta^{\hat{j}\hat{\ell}}
\tilde{W}^j_{\hat{j}}(x) \tilde{W}^{\ell}_{\hat{\ell}}(x).
\end{eqnarray}
One may show that the metric tensors $\tilde{g}_{\mu\nu}(x)$ and
$\tilde{g}^{\mu\nu}(x)$, given by Eq.(\ref{eq:21}) and
Eq.(\ref{eq:30}), respectively, obey the relation
$\tilde{g}^{\mu\alpha}(x) \tilde{g}_{\alpha\nu}(x) =
\delta^{\mu}_{\nu}$. Then, because of $\tilde{e}^{\hat{0}}_j(x) =
\tilde{e}^0_{\hat{j}}(x) = 0$ for the vierbein fields Eq.(\ref{eq:25})
and Eq.(\ref{eq:27}) we get $\tilde{E}^{\hat{0}}_0(x) =
\tilde{e}^{\hat{0}}_0(x)$.

For the vierbein fields Eq.(\ref{eq:25}) and Eq.(\ref{eq:27})
torsion--fermion interactions are yielded by the operators
$C_{\hat{\ell}}$, $G_{\hat{j}}$ and $K$ only. Since the contributions
of the torsion--fermion interactions, caused by the operators
$G_{\hat{j}}$ and $K$, are suppressed by a factor of $1/m$,
below we analyse the contributions of the
  torsion--fermion interactions, caused by the operator
  $C^{\hat{\ell}}$, which appear to order $O(1)$ in the large fermion
  mass expansion. We take also into account the contributions of the
  operators $A$, $B$ and $L^j$ in order to derive a complete set of
  gravitational, chameleon and torsion interactions with slow fermions
  to order $O(1)$ in the large fermion mass expansion. The analysis of
  contributions of the operators $G_{\hat{j}}$ and $K$ goes beyond the
  scope of this paper. We are planning to perform such an analysis in
  our forthcoming publication.

For the vierbein fields Eq.(\ref{eq:25}) and Eq.(\ref{eq:27}) the
operators Eq.(\ref{eq:A.2}) are given by
\begin{eqnarray}\label{eq:31}
A &=& \tilde{e}^{\hat{0}}_0(x),\nonumber\\ B &=& -
\frac{1}{2}\,i\,\tilde{e}^{\hat{0}}_0(x)\,\frac{1}{\sqrt{-
    \tilde{g}}}\,\frac{\partial }{\partial x^j}\Big(\sqrt{-
  \tilde{g}}\,\tilde{e}^j_{\hat{0}}(x)\Big) +
\frac{1}{2}\,i\,(\tilde{e}^{\hat{0}}_0(x))^2\,\tilde{e}^j_{\hat{0}}(x)
\,\frac{1}{\sqrt{- \tilde{g}(x)}}\frac{\partial }{\partial
  x^j}\Big(\sqrt{-
  \tilde{g}(x)}\,\tilde{e}^0_{\hat{0}}(x)\Big),\nonumber\\
 C^{\hat{\ell}}
&=& \frac{1}{4}\,\tilde{e}^{\hat{0}}_0(x)\,\Big(\tilde{\omega}_{0
  \hat{j}\hat{k}}(x)\,\tilde{e}^0_{\hat{0}}(x) + \tilde{\omega}_{\ell
  \hat{j}\hat{k}}(x)\,\tilde{e}^{\ell}_{\hat{0}}(x) +
\tilde{\omega}_{\ell \hat{0}\hat{j}}(x)\,
\tilde{e}^{\ell}_{\hat{k}}(x) - \tilde{\omega}_{\ell \hat{j}
  \hat{0}}(x)\, \tilde{e}^{\ell}_{\hat{k}}(x)\Big)\,
\epsilon^{\hat{j}\hat{k}\hat{\ell}},\nonumber\\ 
D^j_{\hat{j}} &=& -
\tilde{e}^{\hat{0}}_0(x)\,\tilde{e}^j_{\hat{j}}(x),\nonumber\\ F_{\hat{j}}
&=& 0,\nonumber\\
G_{\hat{j}} &=& \frac{1}{2}\,
\tilde{e}^{\hat{0}}_0(x)\,\Big({\tilde{\cal
    T}^{\alpha}\,\!\!}_{\alpha\ell}(x) \,\tilde{e}^{\ell}_{\hat{j}}(x)
+ \tilde{\omega}_{0\hat{j}\hat{0}}(x)\, \tilde{e}^0_{\hat{0}}(x) +
\tilde{\omega}_{\ell\hat{j}\hat{0}}(x)\, \tilde{e}^{\ell}_{\hat{0}}(x)
+ \tilde{\omega}_{\ell\hat{j}\hat{k}}(x)\,
\tilde{e}^{\ell}_{\hat{\ell}}(x)\,\eta^{\hat{\ell}\hat{k}}\Big)\nonumber\\ &+&
\frac{1}{2}\,(\tilde{e}^{\hat{0}}_0(x))^2\,\tilde{e}^j_{\hat{j}}(x)
\,\frac{1}{\sqrt{- \tilde{g}(x)}}\,\frac{\partial }{\partial
  x^j}\Big(\sqrt{-
  \tilde{g}(x)}\,\tilde{e}^0_{\hat{0}}(x)\Big),\nonumber\\ K &=& -
\frac{1}{4}\,\tilde{\omega}_{\ell
  \hat{j}\hat{k}}(x)\,\tilde{e}^{\hat{0}}_0(x)\,
\tilde{e}^{\ell}_{\hat{\ell}}(x)\,\epsilon^{\hat{j}\hat{k}\hat{\ell}},\nonumber\\ L^j
&=& - \tilde{e}^{\hat{0}}_0(x)\,\tilde{e}^j_{\hat{0}}(x).
\end{eqnarray}
For the analysis of interactions of slow Dirac fermions with
gravitational, chameleon and torsion fields, caused by non--diagonal
space--time components $\tilde{e}^j_{\hat{0}}(x)$ of the vierbein
fields we assume a motion of Dirac fermions with mass $m$ in the
curved spacetime with the Schwarzschild metric, taken in the weak
gravitational field of the Earth approximation and modified by the
contributions of the chameleon field and rotation. The line element of
such a spacetime is given by
\begin{eqnarray}\label{eq:32}
d\tilde{s}^2 = (1 + 2 U_+)\,dt^2 + 2\,(1 - 2 U_-)\,\vec{K}\cdot
d\vec{r}\,dt - (1 - 2 U_-)\,d\vec{r}^{\,2},
\end{eqnarray}
where we have neglected the contribution of the terms of order
$\vec{K}^{\,2}$ that is well justified in the terrestrial laboratories
and kept the contributions of the chameleon field to linear order \cite{Hehl1990}. The
potentials $U_{\pm}$ are equal to \cite{Ivanov2015a}
\begin{eqnarray}\label{eq:33}
U_{\pm} = U_{\rm E} \pm \frac{\beta}{M_{\rm Pl}}\,\phi(x),
\end{eqnarray}
where $U_{\rm E} = \vec{g}\cdot \vec{r}$ is the Newtonian
gravitational potential of the Earth and $\vec{g}$ is the
gravitational acceleration \cite{Ivanov2015a}. To linear order
contributions of the gravitational and chameleon field the vierbein
fields Eq.(\ref{eq:25}) and Eq.(\ref{eq:27}) read
\begin{eqnarray}\label{eq:34}
\tilde{e}^{\hat{0}}_0(x) &=&1 + U_+\;,\;\tilde{e}^{\hat{j}}_0(x) = -
(1 - U_-)\,K^{\hat{j}}(x)\;,\; \tilde{e}^{\hat{0}}_j(x) = 0\;,\;
e^{\hat{j}}_j(x) = (1 - U_-)\,
\delta^{\hat{j}}_j,\nonumber\\ \tilde{e}^0_{\hat{0}}(x) &=& 1 -
U_+\;,\; \tilde{e}^j_{\hat{0}}(x) = +(1 - U_+)\, K^j(x)\;,\;
\tilde{e}^0_{\hat{j}}(x) = 0\;,\; \tilde{e}^j_{\hat{j}}(x) = (1 +
U_-)\,\delta^j_{\hat{j}}.
\end{eqnarray}
The diagonal components of the vierbein fields agree well with those,
calculated in \cite{Ivanov2015a}. In such a spacetime the operators
$A$, $B$, $C^{\hat{\ell}}$ and $L^j$ are equal to
\begin{eqnarray}\label{eq:35}
A &=& 1 + U_+,\nonumber\\ B &=& - \frac{1}{2}\,i\,{\rm
  div}\,\vec{K},\nonumber\\ C^{\ell} &=& - \frac{1}{4}\,({\rm
  rot}\,\vec{K}\,)^{\hat{\ell}} + \frac{1}{4}\,{\cal B}^{\ell} +
\frac{1}{4}\,\epsilon^{\ell j k}\,K_j\,{\cal T}_{00k} +
\frac{1}{4}\,\epsilon^{\ell j k}\,{\cal T}_{jka}\,K^a =
\nonumber\\ &=& - \frac{1}{4}\,({\rm rot}\,\vec{K}\,)^{\hat{\ell}} +
\frac{1}{4}\,{\cal B}^{\ell} + \frac{1}{6}\,{\cal K}\,K^{\ell} +
\frac{1}{4}\,\epsilon^{\ell j k}\,K_j\,{\cal M}_{00k} +
\frac{1}{4}\,\epsilon^{\ell j k}\,{\cal M}_{jka}\,K^a,\nonumber\\ L^j
&=& - K^j,
\end{eqnarray}
where to linear order approximation $\tilde{\cal T}_{\alpha\mu\nu} =
{\cal T}_{\alpha\mu\nu}$, ${\cal B}^{\ell} =
\frac{1}{2}\,\epsilon^{\ell jk}\,({\cal T}_{jk0} + {\cal T}_{k0j} +
     {\cal T}_{0jk})$ and ${\cal K} =
     \frac{1}{2}\,\epsilon^{abc}\,{\cal T}_{abc}$. Then, we have used
     Eq.(\ref{eq:6}) and $\sqrt{- \tilde{g}} = 1 + U_+ - 3 U_-$,
     calculated to linear order of the gravitational and chameleon
     field and at the neglect the contribution of order
     $O(\vec{K}^{\,2})$.  For the calculation of the operators in
     Eq.(\ref{eq:35}) we have not distinguished indices in the
     Minkowski and curved spacetime. This is correct, since the
     operators Eq.(\ref{eq:35}) are defined in the perturbative regime
     for gravitational, chameleon and torsion fields and describe
     corresponding interactions of slow Dirac fermions in the
     Minkowski spacetime.

For curved spacetimes with the metric Eq.(\ref{eq:32}) the
contribution of the operators Eq.(\ref{eq:35}) to the effective
low--energy potential $\Phi_{\rm eff}(t,\vec{r},\vec{\sigma}\,)$,
calculated to order $O(1)$ in the large fermion mass expansion, is
given by
\begin{eqnarray}\label{eq:36}
\Phi_{\rm eff}(t,\vec{r},\vec{\sigma}\,) &=& m\,U_+ -
\frac{1}{2}\,i\,{\rm div}\,\vec{K} - \vec{K}\cdot i\,\vec{\nabla} +
\frac{1}{4}\,\vec{\sigma}\cdot {\rm rot}\vec{K} -
\frac{1}{4}\,\vec{\sigma}\cdot \vec{\cal B}\nonumber\\ &-&
\frac{1}{6}\,{\cal K}\,\vec{\sigma}\cdot \vec{K} +
\frac{1}{4}\,\sigma_{\ell}\,\epsilon^{\ell j k}\,K_j\,{\cal M}_{00k} +
\frac{1}{4}\,\sigma_{\ell}\,\epsilon^{\ell jk}\,{\cal M}_{jk a}\,K^a,
\end{eqnarray}
where $\sigma_j = (- \vec{\sigma}\,)_j$ and we have used that
$\epsilon_{jka0} = \epsilon_{jka}$ with $\epsilon_{123} = + 1$
\cite{Itzykson1980}. It is important to note that torsion vector
$\vec{\cal E}$ components do no couple to slow Dirac fermions to order
$O(1)$ in the large fermion mass expansion and to linear order
approximation of the torsion field.

For rotating coordinate systems with an angular velocity
$\vec{\omega}$ the vector functions $K^j$ are equal to $K^j =
-(\vec{\omega}\times \vec{r}\,)^j$ \cite{Hehl1990,LL2008}. This gives
${\rm div}\vec{K} = 0$ and ${\rm rot}\vec{K} = - 2\,\vec{\omega}$. As
a result, the effective low--energy potential Eq.(\ref{eq:36}) takes
the form
\begin{eqnarray}\label{eq:37}
\Phi_{\rm eff}(t,\vec{r},\vec{\sigma}\,) &=& m\,U_+ -
\vec{\omega}\cdot \vec{L} - \vec{\omega}\cdot
\vec{S} - \frac{1}{4}\,\vec{\sigma}\cdot \vec{\cal B} +
\frac{1}{6}\,{\cal K}\,\vec{\sigma}\cdot (\vec{\omega} \times
\vec{r}\,)\nonumber\\ &-& \frac{1}{4}\,\Big((\vec{\sigma}\cdot
\vec{\omega}\,)(\vec{\cal M}\cdot \vec{r}\,) - (\vec{\sigma}\cdot
\vec{r}\,)(\vec{\cal M}\cdot \vec{\omega}\,)\Big) -
\frac{1}{4}\,\sigma_j\,\epsilon^{jk\ell}\,{\cal M}_{k\ell
  a}\,\epsilon^{abc}\omega_b x_c,
\end{eqnarray}
where $(\vec{\cal M}\,)_k = - {\cal M}_{00k}$.  The first term
$m\,U_+$ in Eq.(\ref{eq:36}) corresponds to the Newtonian
gravitational potential of the Earth $\vec{g}\cdot \vec{r}$, corrected
by the contribution of the chameleon field \cite{Brax2011,Ivanov2013},
the second term $ - \vec{\omega}\cdot \vec{L}$, where $\vec{L} = -
\vec{r}\times i\,\vec{\nabla}$ is the orbital momentum operator of
slow fermions, and the third term $ - \vec{\omega}\cdot \vec{S}$,
where $\vec{S} = \frac{1}{2}\,\vec{\sigma}$ is the spin operator of
slow fermions, agree well with the results, obtained by Hehl and Ni
\cite{Hehl1990}. The interactions $- \vec{\omega}\cdot \vec{L}$ and $
- \vec{\omega}\cdot \vec{S}$ were analysed in the experiments by
Werner, Staudenmann, and Colella \cite{Werner1979} and by Mashhoon
\cite{Mashhoon1988}. The fourth term, describing torsion--spin--matter
interaction of the torsion axial--vector components, was derived by
L\"ammerzahl \cite{Laemmerzahl1997} and Obukhov, Silenko, and Teryaev
\cite{Obukhov2014} (see also \cite{Ivanov2015a}). The other terms in
the effective low--energy potential Eq.(\ref{eq:36}) are new. The
fifth term $(1/6)\,{\cal K}\,\vec{\sigma}\cdot (\vec{\omega}\times
\vec{r}\,)$ is a new low--energy interaction of the torsion
pseudoscalar component ${\cal K}$ with slow Dirac fermions. In turn,
the last two terms in Eq.(\ref{eq:36}) describe new low--energy
interactions of torsion tensor ${\cal M}_{00k}$ and ${\cal M}_{jka}$
components with slow Dirac fermions, caused by minimal
torsion--fermion couplings without phenomenological coupling
constants.  According to estimates by Kostelecky {\it et al.}
\cite{Kostelecky2008}, the constant torsion tensor components ${\cal
  M}_{00k}$ and ${\cal M}_{jka}$, multiplied by a phenomenological
coupling constant $\xi^{(5)}_5$, are restricted by $|\xi^{(5)}_5{\cal
  M}_{00k}| < 10^{-27}$ and $|\xi^{(5)}_5{\cal M}_{jka}| < 10^{-26}$,
respectively.  Recently the non--minimal torsion--matter couplings
have been also discussed by Puetzfeld and Obukhov
\cite{Obukhov2014a}. As regards torsion vector components $\vec{\cal
  E}$, we have found that slow Dirac fermions do not couple to them to
order $O(1)$ in the large fermion mass expansion and to linear order
of the torsion field approximation.

The upper bound of the linear superposition of the constant torsion
vector and axial--vector components $|\zeta| < 9.1\times
10^{-23}\,{\rm GeV}$, measured by Lehnert, Snow and Yan
\cite{Lehnert2014} by means of an investigation of a spin rotation of
cold neutrons in the liquid ${^4}{\rm He}$, is by a factor $10^5$
larger compared with the estimate $|\zeta| < 10^{-27}\,{\rm GeV}$,
obtained in \cite{Ivanov2015a}.

Thus, we have shown that to linear order of the torsion field
approximation in spacetimes of rotating coordinate systems the
contributions of only torsion pseudoscalar ${\cal K}$ and tensor
${\cal M}_{00k}$ and $ {\cal M}_{k\ell a}$ components, caused by
minimal torsion--fermion couplings, appear to order $O(1)$ in the
large fermion mass expansion. The certain steps in the realization of
curved spacetimes in terrestrial laboratories by using rotating
devices (neutron interferometers) were made by Atwood {\it et al.}
\cite{Atwood1984} and Mashhoon \cite{Mashhoon1988}.

The measurements of the transition frequencies between quantum
gravitational states of ultracold neutrons in the qBounce experiments
\cite{Abele2010}--\cite{Jenke2014} as functions of an angular velocity
$\vec{\omega}$ of a rotating mirror should provide a new level of
highly precise probes of the properties of the Einstein--Cartan
gravity, dark energy and evolution of the Universe. Of course, the
measurements of new gravitational, chameleon and torsion interactions
in Eq.(\ref{eq:37}) as well as other interactions in the effective
low--energy potential Eq.(\ref{eq:A.15}) can be carried out by using
rotating neutron interferometers
\cite{Lemmel2015,Rauch2015}.

Now we would like to discuss shortly the Foldy--Wouthuysen method
\cite{Foldy1950}, which we use in this paper for the derivation of the
effective low--energy potential for slow Dirac fermions, coupled to
gravitational, chameleon and torsion fields. Mainly this discussion
concerns uniqueness and accuracy of the Foldy--Wouthuysen
representation of the Dirac Hamilton operator, obtained by the
Foldy--Wouthuysen transformation. It is well--known that the
Foldy--Wouthuysen method of a transformation of a Dirac Hamilton
operator to a form, containing only {\it even} (diagonal) operators
(as regards the definition of {\it odd} and {\it even} operators a
reader might consult Ref.\cite{Foldy1950} or look up in the Appendix
to this paper), is not unique and there are some other methods of
transformation of a Dirac Hamilton operator to a diagonal form. A very
nice survey of possible methods of transformation of a Dirac Hamilton
operator for fermions with mass $m$ to a diagonal form, containing
only {\it even} operators, one can find in the paper by Vries
\cite{Vries1970}. The Foldy--Wouthuysen method, removing {\it odd}
operators from a Dirac Hamilton operator for fermions with mass $m$ by
Foldy--Wouthuysen unitary transformations, allows to reduce a Dirac
Hamilton operator to a non--relativistic form in the approximation of
a large fermion mass expansion by a set of unitary transformations or
by the iterative Foldy--Wouthuysen method.  The obtained
non--relativistic Hamilton operator is given by an infinite series of
{\it even} operators in powers of $1/m$, which does not seem to give
hope for a closed--form operator. A problem of a closed form of a
transformed Dirac Hamilton operator, expressed in terms of only {\it
  even} operators, was investigated by Eriksen
\cite{Eriksen1958}. Eriksen showed that the unitary transformation
$e^{\,iS} = \sqrt{\gamma^0 {\rm H}/\sqrt{{\rm H}^2}}$, where
$\gamma^0$ and ${\rm H}$ are the Dirac matrix and a Dirac Hamilton
operator, allows to transform a Dirac Hamilton operator ${\rm H}$ to a
square root of an {\it even} operator. However, Eriksen's unitary
operator $e^{\,iS} = \sqrt{\gamma^0 {\rm H}/\sqrt{{\rm H}^2}}$,
leading to a closed--form of a transformed Dirac Hamilton operators,
suffers from ambiguous definition. In order to define the operator
$e^{\,iS} = \sqrt{\gamma^0 {\rm H}/\sqrt{{\rm H}^2}}$ unambiguously
one has to assume that the square root of a unit operator is a unit
operator. For recent discussion of a square root operator definition
and analyses of the Dirac Hamilton operators by means of the Eriksen
method we propose a reader the papers by Silenko \cite{Silenko2003},
Neznamov and Silenko \cite{Neznamov2009} and Silenko
\cite{Silenko2013,Silenko2015,Silenko2015a}. According
  to Eriksen \cite{Eriksen1958}, the unitary operator $e^{\,iS} =
  \sqrt{\gamma^0 {\rm H}/\sqrt{{\rm H}^2}}$, providing an {\it exact}
  diagonalization of the Dirac Hamilton operator ${\rm H}$, can be
  defined by $e^{\,iS} = \sqrt{\gamma^0 \lambda} = (1 + \gamma^0
  \lambda)/\sqrt{2 + \gamma^0 \lambda + \lambda \gamma^0}$, where
  $\lambda = {\rm H}/\sqrt{{\rm H}^2}.$ Another problem of the
Foldy--Wouthuysen method concerns an accuracy of the Foldy--Wouthuysen
representation of a Dirac Hamilton operator in comparison with a large
fermion mass expansion of an {\it exact} form of a transformed Dirac
Hamilton operator. For the first time such a problem was discussed by
Eriksen and Kolsrud \cite{Eriksen1960}.  Recently this problem has
been investigated by Neznamov and Silenko \cite{Neznamov2009} and
Silenko \cite{Silenko2013,Silenko2015,Silenko2015a}. According to
\cite{Eriksen1960} and
\cite{Neznamov2009,Silenko2013,Silenko2015,Silenko2015a}, the
Foldy--Wouthuysen representation of a Dirac Hamilton operator,
obtained by a set of unitary transformations, can but not coincide
with a large fermion mass expansion of an {\it exact} transformed
Dirac Hamilton operator, diagonalized by means of only one unitary
transformation (e.g. the Eriksen transformation). As has been pointed
out by Neznamov and Silenko \cite{Neznamov2009} and Silenko
\cite{Silenko2013,Silenko2015,Silenko2015a}, such a disagreement can
be explained by a non--commutativity of Foldy--Wouthuysen unitary
transformations in the iterative Foldy--Wouthuysen method. For
example, in our case we have diagonalized the Dirac Hamilton operator
Eq.(\ref{eq:17}) by three unitary transformations of the Dirac wave
functions $e^{\,iS_1}$, $e^{\,iS_2}$ and $e^{\,iS_3}$ (see the
Appendix), respectively. A resulting unitary transformation is equal
to $e^{i\,S} = e^{\,iS_3}e^{i\,S_2}e^{i\,S_1}$. According to
\cite{Neznamov2009,Silenko2013,Silenko2015,Silenko2015a}, a
coincidence of our result for the effective low--energy potential
Eq.(\ref{eq:A.15}) with a large fermion mass expansion of an {\it
  exact} diagonalized Hamilton operator, obtained by means of only one
unitary transformation (e.g. by the Eriksen transformation), can be
expected only for the validity of the relation
$e^{i\,S_3}e^{\,iS_2}e^{i\,S_1} = e^{\,i\,(S_3 + S_2 + S_1)}$, which
demands a commutativity of the operators $[S_i,S_j] = 0$ for $i\neq j
= 1,2,3$
\cite{Neznamov2009,Silenko2013,Silenko2015,Silenko2015a}. Since the
unitary operators do not commute $[S_i,S_j] \neq 0$ for $i\neq j =
1,2,3$ (see Appendix), such the relation
$e^{i\,S_3}e^{\,iS_2}e^{i\,S_1} = e^{\,i\,(S_3 + S_2 + S_1)}$ is not
valid and one may expect some deviations of the effective low--energy
potential Eq.(\ref{eq:A.15}) from that derived by a large fermions
mass expansion of an {\it exact} diagonalized Dirac Hamilton
operator. However, one may show that any deviations can appear only to
order $O(1/m^2)$ \cite{Silenko2015a}. The later can be justified by
the observation that $[S_2,S_1] = O(1/m^2)$ and $[S_3, S_j] =
O(1/m^3)$ for $j = 1,2$. Hence, to order $O(1/m)$, which we have kept
for the derivation of the effective low--energy potential $\Phi_{\rm
  eff}$ in Eq.(\ref{eq:A.15}), these two fermion mass expansions
should coincide. A method of the calculation of the corrections to the
Foldy--Wouthuysen representation of a Dirac Hamilton operator has been
discussed in detail by Silenko \cite{Silenko2015a}. For example,
suppose that two Foldy--Wouthuysen unitary transformations with
operators $e^{\,iS_1}$ and $e^{\,iS_2}$, performed one after another,
diagonalize a Dirac Hamilton operator, i.e. ${\rm H} \to {\rm H}_{\rm
  FW}$, where ${\rm H}_{\rm FW}$ is a Dirac Hamilton operator in the
Foldy--Wouthuysen representation. According to Silenko
\cite{Silenko2015a}, an additional Foldy--Wouthuysen unitary
transformation $U_{\rm corr} = \exp(- \frac{1}{2}\,[S_1,S_2])$ should
allow to cancel an error of the iterative Foldy--Wouthuysen method in
the leading order. Such a correction is valid if the commutators
$[S_1,[S_1,S_2]]$ and $[S_2,[S_1,S_2]]$ and commutators of higher
orders can be neglected with respect to the commutator
$[S_1,S_2]$. Since in our case this constraint is fulfilled, the
correction to the effective low--energy potential Eq.(\ref{eq:A.15})
can be calculated by means of the unitary transformation $U_{\rm corr}
= \exp(- \frac{1}{2}\,[S_1,S_2])$. However, in our case $[S_1,S_2] =
O(1/m^2)$ and the effective low--energy potential Eq.(\ref{eq:A.15})
is calculated to order $O(1/m)$. This might imply that the effective
low--energy potential Eq.(\ref{eq:A.15}) should in principle coincide
with a large fermion mass expansion to order $O(1/m)$ of an {\it
  exact} diagonalized Dirac Hamilton operator by, for example, the
Eriksen method \cite{Eriksen1958,
  Neznamov2009,Silenko2013,Silenko2015,Silenko2015a}.

\section{Acknowledgements}

We are grateful to Hartmut Abele for stimulating discussions. This
work was supported by the Austrian ``Fonds zur F\"orderung der
Wissenschaftlichen Forschung'' (FWF) under the contract I689-N16.

\section{Appendix A: Derivation of the effective low--energy potential 
$\Phi_{\rm eff}(t,\vec{r},\vec{\sigma}\,)$ in Eq.(\ref{eq:18})}
\renewcommand{\theequation}{A-\arabic{equation}}
\setcounter{equation}{0}

For the derivation of the effective low--energy potential $\Phi_{\rm
  eff}(t,\vec{r},\vec{\sigma}\,)$ for slow Dirac fermions with mass
$m$ in the Sch\"odinger--Pauli equation Eq.(\ref{eq:18}) we define the
Hamilton operator Eq.(\ref{eq:17}) as follows
\begin{eqnarray}\label{eq:A.1}
{\rm H}' = A\gamma^{\hat{0}}m + B + C^{\hat{\ell}}\Sigma_{\hat{\ell}}
+
D^j_{\hat{j}}i\gamma^{\hat{0}}\gamma^{\hat{j}}\frac{\partial}{\partial
  x^j} +
F_{\hat{j}}i\gamma^{\hat{0}}\gamma^{\hat{j}}\frac{\partial}{\partial
  t} + G_{\hat{j}}i\gamma^{\hat{0}}\gamma^{\hat{j}} + K\,\gamma^5 +
L^j\,i\,\frac{\partial}{\partial x^j}
\end{eqnarray}
where we have denoted
\begin{eqnarray}\label{eq:A.2}
A &=& \tilde{E}^{\hat{0}}_0(x),\nonumber\\ 
B &=& -
\frac{1}{2}\,i\,\tilde{E}^{\hat{0}}_0(x)\,\frac{1}{\sqrt{-
    \tilde{g}}}\,\frac{\partial }{\partial x^j}\Big(\sqrt{-
  \tilde{g}}\,\tilde{e}^j_{\hat{0}}(x)\Big) +
\frac{1}{2}\,i\,(\tilde{E}^{\hat{0}}_0(x))^2\,\tilde{e}^j_{\hat{0}}(x)
\,\frac{1}{\sqrt{- \tilde{g}(x)}}\frac{\partial }{\partial
  x^j}\Big(\sqrt{-
  \tilde{g}(x)}\,\tilde{e}^0_{\hat{0}}(x)\Big),\nonumber\\ 
C^{\hat{\ell}}
&=& \frac{1}{4}\,\tilde{E}^{\hat{0}}_0(x)\,\Big(\tilde{\omega}_{\mu
  \hat{j}\hat{k}}(x)\,\tilde{e}^{\mu}_{\hat{0}}(x) +
\tilde{\omega}_{\mu [\hat{0}\hat{j}]}(x)\,
\tilde{e}^{\mu}_{\hat{k}}(x)\Big)\,
\epsilon^{\hat{j}\hat{k}\hat{\ell}},\nonumber\\ 
D^j_{\hat{j}} &=& -
\tilde{E}^{\hat{0}}_0(x)\,\tilde{e}^j_{\hat{j}}(x),\nonumber\\ 
F_{\hat{j}}
&=& -
\tilde{E}^{\hat{0}}_0(x)\,\tilde{e}^0_{\hat{j}}(x),\nonumber\\ 
G_{\hat{j}}
&=& \frac{1}{2}\, \tilde{E}^{\hat{0}}_0(x)\,\Big({\tilde{\cal
    T}^{\alpha}\,\!\!}_{\alpha\mu}(x) \tilde{e}^{\mu}_{\hat{j}}(x) +
\tilde{\omega}_{\mu\hat{j}\hat{\beta}}(x)
\tilde{e}^{\mu}_{\hat{\lambda}}(x)\,\eta^{\hat{\lambda}\hat{\beta}}\Big)
- \frac{1}{2}\,(\tilde{E}^{\hat{0}}_0(x))^2\,\tilde{e}^0_{\hat{j}}(x)
\,\Big({{\cal \tilde{T}}^{\alpha}\,}_{\alpha\mu}(x)
\tilde{e}^{\mu}_{\hat{0}}(x) +
\tilde{\omega}_{\mu\hat{0}\hat{\beta}}(x)\tilde{e}^{\mu}_{\hat{\lambda}}(x)\,
\eta^{\hat{\lambda}\hat{\beta}}\Big)\nonumber\\ 
&+&
\frac{1}{2}\,(\tilde{E}^{\hat{0}}_0(x))^2\,\tilde{e}^j_{\hat{j}}(x)
\,\frac{1}{\sqrt{- \tilde{g}(x)}}\,\frac{\partial }{\partial
  x^j}\Big(\sqrt{-
  \tilde{g}(x)}\,\tilde{e}^0_{\hat{0}}(x)\Big) -
\frac{1}{2}\,(\tilde{E}^{\hat{0}}_0(x))^2\,\tilde{e}^0_{\hat{j}}(x)
\,\frac{1}{\sqrt{- \tilde{g}(x)}}\,\frac{\partial }{\partial
  x^j}\Big(\sqrt{-
  \tilde{g}(x)}\,\tilde{e}^j_{\hat{0}}(x)\Big),\nonumber\\ 
K &=& -
\frac{1}{4}\,\tilde{E}^{\hat{0}}_0(x)\,\tilde{\omega}_{\mu
  \hat{j}\hat{k}}(x)\,
\tilde{e}^{\mu}_{\hat{\ell}}(x)\,\epsilon^{\hat{j}\hat{k}\hat{\ell}},\nonumber\\ L^j
&=& - \tilde{E}^{\hat{0}}_0(x)\,\tilde{e}^j_{\hat{0}}(x).
\end{eqnarray}
For the elimination of the {\it odd} operators we perform the
Foldy--Wouthuysen unitary transformation of the wave function
$\psi'(x) = e^{\,-i S_1}\,\psi_1(x)$ and the Hamilton operator
\cite{Foldy1950}
\begin{eqnarray}\label{eq:A.3}
{\rm H}_1 = e^{\,+ i S_1}\,{\rm H}'\,e^{\,-i S_1} - i\,e^{\,i
  S_1}\frac{\partial}{\partial t}e^{\,-i S_1} = {\rm H}' -
\frac{\partial S_1}{\partial t} + i\Big[S_1,{\rm H}' -
  \frac{1}{2}\,\frac{\partial S_1}{\partial t}\Big] +
\frac{i^2}{2}\,\Big[S_1,\Big[S_1,{\rm H}' -
    \frac{1}{3}\,\frac{\partial S_1}{\partial t}\Big]\Big] + \ldots
\end{eqnarray}
The time derivative appears because of a time dependence of the
chameleon and torsion fields. Then, following \cite{Foldy1950} we take
the operator $S_1$ in the form
\begin{eqnarray}\label{eq:A.4}
S_1 =  - \frac{i}{2m A}\,\gamma^{\hat{0}}\Big(
D^j_{\hat{j}}\,i\gamma^{\hat{0}}\gamma^{\hat{j}}\frac{\partial}{\partial
  x^j} +
F_{\hat{j}}\,i\gamma^{\hat{0}}\gamma^{\hat{j}}\frac{\partial}{\partial
  t} + G_{\hat{j}}\,i\gamma^{\hat{0}}\gamma^{\hat{j}} +
K\,\gamma^5\Big).
\end{eqnarray}
The time derivative of $S_1$ and the commutators in Eq.(\ref{eq:A.3})
are equal to
\begin{eqnarray}\label{eq:A.5}
\frac{\partial S_1}{\partial t} =
\frac{1}{2m}\,\frac{\partial}{\partial
  t}\Big(\frac{D^j_{\hat{j}}}{A}\Big)\,
\gamma^{\hat{j}}\frac{\partial}{\partial x^j} +
\frac{1}{2m}\,\frac{\partial}{\partial
  t}\Big(\frac{F_{\hat{j}}}{A}\Big)\,\gamma^{\hat{j}}\frac{\partial}{\partial
  t} + \frac{1}{2m}\,\frac{\partial}{\partial
  t}\Big(\frac{G_{\hat{j}}}{A}\Big)\,\gamma^{\hat{j}} -
\frac{1}{2m}\,\frac{\partial }{\partial
  t}\Big(\frac{K}{A}\Big)\,i\,\gamma^{\hat{0}}\gamma^5
\end{eqnarray}
and
\begin{eqnarray*}
&&i\Big[S_1,{\rm H}' - \frac{1}{2}\,\frac{\partial S_1}{\partial
      t}\Big] =\nonumber\\ &&= -
  D^j_{\hat{j}}i\gamma^{\hat{0}}\gamma^{\hat{j}}\frac{\partial}{\partial
    x^j} -
  F_{\hat{j}}i\gamma^{\hat{0}}\gamma^{\hat{j}}\frac{\partial}{\partial
    t} - G_{\hat{j}}i\gamma^{\hat{0}}\gamma^{\hat{j}} - K\,\gamma^5 -
  \frac{1}{2}\,i\,\gamma^{\hat{0}}\gamma^{\hat{j}}\frac{D^j_{\hat{j}}}{A}\,\frac{\partial
    A}{\partial x^j} -
  \frac{1}{2}\,i\,\gamma^{\hat{0}}\gamma^{\hat{j}}\frac{F_{\hat{j}}}{A}\,\frac{\partial
    A}{\partial t}\nonumber\\ &&+ \frac{1}{2m
    A}\,i\,\gamma^{\hat{j}}\, D^j_{\hat{j}}\,\frac{\partial
    B}{\partial x^j} + \frac{1}{2m A}\,i\,\gamma^{\hat{j}}\,
  F_{\hat{j}}\,\frac{\partial B}{\partial t} - \frac{1}{2m
    A}\,\eta^{\hat{j}\hat{\ell}}D^j_{\hat{j}}\frac{\partial
    C_{\hat{\ell}}}{\partial x^j}\,i\,\gamma^{\hat{0}}\gamma^5 +
  \frac{1}{2m
    A}\,\epsilon^{\hat{j}\hat{\ell}\hat{k}}\,\gamma_{\hat{k}}\,D^j_{\hat{j}}\,
  \frac{\partial C_{\hat{\ell}}}{\partial x^j}\nonumber\\ &&+
  \frac{1}{m
    A}\,\epsilon^{\hat{j}\hat{\ell}\hat{k}}\,\gamma_{\hat{k}}\,D^j_{\hat{j}}
  C_{\hat{\ell}}\,\frac{\partial }{\partial x^j} - \frac{1}{2m
    A}\,\eta^{\hat{j}\hat{\ell}}F_{\hat{j}}\frac{\partial
    C_{\hat{\ell}}}{\partial t}\,i\,\gamma^{\hat{0}}\gamma^5 +
  \frac{1}{2m
    A}\,\epsilon^{\hat{j}\hat{\ell}\hat{k}}\,\gamma_{\hat{k}}\,F_{\hat{j}}\,
  \frac{\partial C_{\hat{\ell}}}{\partial t}\nonumber\\
\end{eqnarray*}
\begin{eqnarray}\label{eq:A.6}
&&+ \frac{1}{m
    A}\,\epsilon^{\hat{j}\hat{\ell}\hat{k}}\,\gamma_{\hat{k}}\,F_{\hat{j}}
  C_{\hat{\ell}}\,\frac{\partial }{\partial t} + \frac{1}{m
    A}\,\epsilon^{\hat{j}\hat{\ell}\hat{k}}\,\gamma_{\hat{k}}\,G_{\hat{j}}\,
  C_{\hat{\ell}} + \frac{1}{m
    A}\,\gamma^{\hat{0}}\,\eta^{\hat{j}\hat{k}}\,D^j_{\hat{j}}D^k_{\hat{k}}\,
  \frac{\partial^2}{\partial x^j \partial x^k}\nonumber\\ 
&&+
  \frac{1}{2 m
    A}\,\gamma^{\hat{0}}\,\eta^{\hat{j}\hat{k}}\,D^j_{\hat{j}}\frac{\partial
    D^k_{\hat{k}}}{\partial x^j}\,\frac{\partial}{\partial x^k} +
  \frac{1}{2m}\,\gamma^{\hat{0}}\,\eta^{\hat{j}\hat{k}}\,D^k_{\hat{k}}\,
  \frac{\partial }{\partial
    x^k}\Big(\frac{D^j_{\hat{j}}}{A}\Big)\,\frac{\partial}{\partial
    x^j} + \frac{1}{2 m
    A}\,\gamma^{\hat{0}}\,i\,\epsilon^{\hat{j}\hat{k}\hat{\ell}}\,
  \Sigma_{\hat{\ell}}\,D^j_{\hat{j}}\frac{\partial
    D^k_{\hat{k}}}{\partial x^j}\,\frac{\partial}{\partial x^k}
  \nonumber\\ 
&&-
  \frac{1}{2m}\,\gamma^{\hat{0}}\,i\,\epsilon^{\hat{j}\hat{k}\hat{\ell}}\,
  \Sigma_{\hat{\ell}}\,D^k_{\hat{k}}\, \frac{\partial }{\partial
    x^k}\Big(\frac{D^j_{\hat{j}}}{A}\Big)\,\frac{\partial}{\partial
    x^j} + \frac{1}{m
    A}\,\gamma^{\hat{0}}\,\eta^{\hat{j}\hat{k}}\,F_{\hat{j}}
  D^k_{\hat{k}}\,\frac{\partial^2}{\partial x^k \partial t} +
  \frac{1}{2 m
    A}\,\gamma^{\hat{0}}\,\eta^{\hat{j}\hat{k}}\,F_{\hat{j}}\frac{\partial
    D^k_{\hat{k}}}{\partial t}\,\frac{\partial}{\partial
    x^k}\nonumber\\ && +
  \frac{1}{2m}\,\gamma^{\hat{0}}\,\eta^{\hat{j}\hat{k}}\,D^k_{\hat{k}}
  \frac{\partial}{\partial x^k}
  \Big(\frac{F_{\hat{j}}}{A}\Big)\,\frac{\partial }{\partial t} +
  \frac{1}{2 m
    A}\,\gamma^{\hat{0}}\,i\,\epsilon^{\hat{j}\hat{k}\hat{\ell}}\,\Sigma_{\hat{\ell}}\,F_{\hat{j}}\,\frac{\partial
    D^k_{\hat{k}}}{\partial t}\,\frac{\partial}{\partial
    x^k}\nonumber\\ 
&& - \frac{1}{2
    m}\,\gamma^{\hat{0}}\,i\,\epsilon^{\hat{j}\hat{k}\hat{\ell}}\,
  \Sigma_{\hat{\ell}}\,D^k_{\hat{k}}\,\frac{\partial}{\partial
    x^k}\Big(\frac{F_{\hat{j}}}{A}\Big)\,\frac{\partial }{\partial t}
  + \frac{1}{m
    A}\,\gamma^{\hat{0}}\,\eta^{\hat{j}\hat{k}}\,G_{\hat{j}}\,D^k_{\hat{k}}\,
  \frac{\partial}{\partial x^k} + \frac{1}{2 m
  }\,\gamma^{\hat{0}}\,\eta^{\hat{j}\hat{k}}\,D^k_{\hat{k}}\,
  \frac{\partial}{\partial
    x^k}\Big(\frac{G_{\hat{j}}}{A}\Big)\nonumber\\ 
&& - \frac{1}{2 m
  }\,\gamma^{\hat{0}}\,i\,\epsilon^{\hat{j}\hat{k}\hat{\ell}}\,
  \Sigma_{\hat{\ell}}\,D^k_{\hat{k}}\,\frac{\partial}{\partial
    x^k}\Big(\frac{G_{\hat{j}}}{A}\Big) + \frac{1}{m
    A}\,i\,\gamma^{\hat{0}}\Sigma^{\hat{k}}\,K\,
  D^k_{\hat{k}}\,\frac{\partial }{\partial x^k} +
  \frac{1}{2m}\,i\,\gamma^{\hat{0}}\Sigma^{\hat{k}}\,D^k_{\hat{k}}\,
  \frac{\partial}{\partial x^k}\Big(\frac{K}{A}\Big)\nonumber\\ 
&& +
  \frac{1}{m
    A}\,\gamma^{\hat{0}}\,\eta^{\hat{j}\hat{k}}\,D^j_{\hat{j}}\,F_{\hat{k}}\,
  \frac{\partial^2}{\partial x^j\partial t} + \frac{1}{2 m
    A}\,\gamma^{\hat{0}}\,\eta^{\hat{j}\hat{k}}\,D^j_{\hat{j}}\,\frac{\partial
    F_{\hat{k}}}{\partial x^j}\,\frac{\partial}{\partial t} +
  \frac{1}{2
    m}\,\gamma^{\hat{0}}\,\eta^{\hat{j}\hat{k}}\,F_{\hat{k}}\,
  \frac{\partial}{\partial
    t}\Big(\frac{D^j_{\hat{j}}}{A}\Big)\,\frac{\partial }{\partial
    x^j}\nonumber\\ 
&& + \frac{1}{2 m
    A}\,\gamma^{\hat{0}}\,i\,\epsilon^{\hat{j}\hat{k}\hat{\ell}}\,
  \Sigma_{\hat{\ell}}\,D^j_{\hat{j}}\,\frac{\partial
    F_{\hat{k}}}{\partial x^j}\,\frac{\partial}{\partial t} -
  \frac{1}{2
    m}\,\gamma^{\hat{0}}\,i\,\epsilon^{\hat{j}\hat{k}\hat{\ell}}\,
  \Sigma_{\hat{\ell}}\,F_{\hat{k}}\,\frac{\partial}{\partial
    t}\Big(\frac{D^j_{\hat{j}}}{A}\Big)\,\frac{\partial }{\partial
    x^j}+ \frac{1}{m
    A}\,\gamma^{\hat{0}}\,\eta^{\hat{j}\hat{k}}\,F_{\hat{j}}\,F_{\hat{k}}\,
  \frac{\partial^2}{\partial t^2}\nonumber\\ 
&& + \frac{1}{2 m
    A}\,\gamma^{\hat{0}}\,\eta^{\hat{j}\hat{k}}\,F_{\hat{j}}\,\frac{\partial
    F_{\hat{k}}}{\partial t}\,\frac{\partial}{\partial t} + \frac{1}{2
    m
  }\,\gamma^{\hat{0}}\,\eta^{\hat{j}\hat{k}}\,F_{\hat{k}}\,\frac{\partial
  }{\partial
    t}\Big(\frac{F_{\hat{j}}}{A}\Big)\,\frac{\partial}{\partial t} +
  \frac{1}{2 m
    A}\,\gamma^{\hat{0}}\,i\,\epsilon^{\hat{j}\hat{k}\hat{\ell}}\,
  \Sigma_{\hat{\ell}}\,F_{\hat{j}}\,\frac{\partial
    F_{\hat{k}}}{\partial t}\,\frac{\partial}{\partial
    t}\nonumber\\ 
&& - \frac{1}{2 m
  }\,\gamma^{\hat{0}}\,i\,\epsilon^{\hat{j}\hat{k}\hat{\ell}}\,
  \Sigma_{\hat{\ell}}\,F_{\hat{k}}\,\frac{\partial }{\partial
    t}\Big(\frac{F_{\hat{j}}}{A}\Big)\,\frac{\partial}{\partial t} +
  \frac{1}{mA}\,\gamma^{\hat{0}}\,\eta^{\hat{j}\hat{k}}\,
  G_{\hat{j}}\,F_{\hat{k}}\, \frac{\partial}{\partial t} + \frac{1}{2
    m
  }\,\gamma^{\hat{0}}\,\eta^{\hat{j}\hat{k}}\,F_{\hat{k}}\,\frac{\partial
  }{\partial t}\Big(\frac{G_{\hat{j}}}{A}\Big)\nonumber\\ 
&& -
  \frac{1}{2 m }\,\gamma^{\hat{0}}\,
  i\,\epsilon^{\hat{j}\hat{k}\hat{\ell}}\,
  \Sigma_{\hat{\ell}}\,F_{\hat{k}}\,\frac{\partial }{\partial
    t}\Big(\frac{G_{\hat{j}}}{A}\Big) +
  \frac{1}{mA}\,i\,\gamma^{\hat{0}}\,\Sigma^{\hat{k}}\,K\,
  F_{\hat{k}}\, \frac{\partial}{\partial t} + \frac{1}{2 m
  }\,i\,\gamma^{\hat{0}}\,\Sigma^{\hat{k}}\,F_{\hat{k}}\,\frac{\partial}{\partial
    t}\Big(\frac{K}{A}\Big)\nonumber\\ 
&& + \frac{1}{m
    A}\,\gamma^{\hat{0}}\,\eta^{\hat{j}\hat{k}}\,G_{\hat{k}}\,D^j_{\hat{j}}
  \frac{\partial}{\partial x^j} + \frac{1}{2 m
    A}\,\gamma^{\hat{0}}\,\eta^{\hat{j}\hat{k}}\,D^j_{\hat{j}}\frac{\partial
    G_{\hat{k}}}{\partial x^j} + \frac{1}{2 m
    A}\,\gamma^{\hat{0}}\,i\,\epsilon^{\hat{j}\hat{k}\hat{\ell}}\,
  \Sigma_{\hat{\ell}}\,D^j_{\hat{j}}\frac{\partial
    G_{\hat{k}}}{\partial x^j}\nonumber\\
&& + \frac{1}{m
    A}\,\gamma^{\hat{0}}\,\eta^{\hat{j}\hat{k}}\,F_{\hat{j}}\,G_{\hat{k}}\,
  \frac{\partial }{\partial t} + \frac{1}{2 m
    A}\,\gamma^{\hat{0}}\,\eta^{\hat{j}\hat{k}}\,F_{\hat{j}}\,\frac{\partial
    G_{\hat{k}} }{\partial t} + \frac{1}{2 m
    A}\,\gamma^{\hat{0}}\,i\,\epsilon^{\hat{j}\hat{k}\hat{\ell}}\,
  \Sigma_{\hat{\ell}}\,F_{\hat{j}}\,\frac{\partial G_{\hat{k}}
  }{\partial t}\nonumber\\
&& + \frac{1}{m
    A}\,\gamma^{\hat{0}}\,\eta^{\hat{j}\hat{k}}\,G_{\hat{j}}\,G_{\hat{k}}
  + \frac{1}{m A}\,i\,\gamma^{\hat{0}}\,\Sigma^{\hat{k}}
  K\,G_{\hat{k}} + \frac{1}{m
    A}\,i\,\gamma^{\hat{0}}\,\Sigma^{\hat{j}}\,K\,D^j_{\hat{j}}\,
  \frac{\partial}{\partial x^j} + \frac{1}{2 m
    A}\,i\,\gamma^{\hat{0}}\,\Sigma^{\hat{j}}\,D^j_{\hat{j}}\,
  \frac{\partial K}{\partial x^j}\nonumber\\
&& + \frac{1}{m
    A}\,i\,\gamma^{\hat{0}}\,\Sigma^{\hat{j}}\,K\,F_{\hat{j}}\,
  \frac{\partial}{\partial t} + \frac{1}{2 m
    A}\,i\,\gamma^{\hat{0}}\,\Sigma^{\hat{j}}\,F_{\hat{j}}\,
  \frac{\partial K}{\partial t}+ \frac{1}{m
    A}\,i\,\gamma^{\hat{0}}\,\Sigma^{\hat{j}}\,K\,G_{\hat{j}} +
  \frac{1}{m A}\,\gamma^{\hat{0}}\,K^2\nonumber\\ 
&& - \frac{1}{2 m
    A}\,\gamma^{\hat{j}}\,D^j_{\hat{j}}\,\frac{\partial L^k}{\partial
    x^j}\,\frac{\partial}{\partial x^k} + \frac{1}{2
    m}\,\,\gamma^{\hat{j}}\,L^k\,\frac{\partial}{\partial
    x^k}\Big(\frac{D^j_{\hat{j}}}{A}\Big)\,\frac{\partial }{\partial
    x^j} - \frac{1}{2 m
    A}\,\gamma^{\hat{j}}\,F_{\hat{j}}\,\frac{\partial L^k}{\partial
    t}\,\frac{\partial}{\partial x^k}\nonumber\\
&& + \frac{1}{2
    m}\,\,\gamma^{\hat{j}}\,L^k\,\frac{\partial}{\partial
    x^k}\Big(\frac{F_{\hat{j}}}{A}\Big)\,\frac{\partial }{\partial t}
  + \frac{1}{2 m}\,\gamma^{\hat{j}}\,L^k\,\frac{\partial}{\partial
    x^k}\Big(\frac{G_{\hat{j}}}{A}\Big) - \frac{1}{2
    m}\,i\,\gamma^{\hat{0}}\,\gamma^5\,L^k\,\frac{\partial}{\partial
    x^k}\Big(\frac{K}{A}\Big)
\end{eqnarray}
and 
\begin{eqnarray*}
&&\frac{i^2}{2}\,\Big[S_1,\Big[S_1,{\rm H}' -
      \frac{1}{3}\,\frac{\partial S_1}{\partial t}\Big]\Big]
  =\nonumber\\ &&= - \frac{1}{2 m
    A}\,\gamma^{\hat{0}}\,\eta^{\hat{j}\hat{k}}\,D^j_{\hat{j}}
  D^k_{\hat{k}}\, \frac{\partial^2 }{\partial x^j \partial x^k} -
  \frac{1}{4 m
    A}\,\gamma^{\hat{0}}\,\eta^{\hat{j}\hat{k}}\,D^j_{\hat{j}}\,\frac{\partial
    D^k_{\hat{k}}}{\partial x^j}\,\frac{\partial}{\partial x^k} -
  \frac{1}{4 m }\,\gamma^{\hat{0}}\,\eta^{\hat{j}\hat{k}}\,
  D^k_{\hat{k}}\,\frac{\partial }{\partial
    x^k}\Big(\frac{D^j_{\hat{j}}}{A}\Big)\,\frac{\partial}{\partial
    x^j}\nonumber\\&& - \frac{1}{4 m
    A}\,\gamma^{\hat{0}}\,i\,\epsilon^{\hat{j}\hat{k}\hat{\ell}}\,
  \Sigma_{\hat{\ell}}\,D^j_{\hat{j}}\,\frac{\partial
    D^k_{\hat{k}}}{\partial x^j}\,\frac{\partial}{\partial x^k} +
  \frac{1}{4 m
  }\,\gamma^{\hat{0}}\,i\,\epsilon^{\hat{j}\hat{k}\hat{\ell}}\,
  \Sigma_{\hat{\ell}}\, D^k_{\hat{k}}\,\frac{\partial }{\partial
    x^k}\Big(\frac{D^j_{\hat{j}}}{A}\Big)\,\frac{\partial}{\partial
    x^j} - \frac{1}{2 m
    A}\,\gamma^{\hat{0}}\,\eta^{\hat{j}\hat{k}}\,F_{\hat{j}}
  D^k_{\hat{k}}\, \frac{\partial^2 }{\partial t\partial x^k}
  \nonumber\\&& - \frac{1}{4 m
    A}\,\gamma^{\hat{0}}\,\eta^{\hat{j}\hat{k}}\,F_{\hat{j}}\,\frac{\partial
    D^k_{\hat{k}}}{\partial t}\,\frac{\partial}{\partial x^k} -
  \frac{1}{4 m }\,\gamma^{\hat{0}}\,\eta^{\hat{j}\hat{k}}\,
  D^k_{\hat{k}}\,\frac{\partial}{\partial
    x^k}\Big(\frac{F_{\hat{j}}}{A}\Big)\,\frac{\partial}{\partial t} -
  \frac{1}{4 m
    A}\,\gamma^{\hat{0}}\,i\,\epsilon^{\hat{j}\hat{k}\hat{\ell}}\,
  \Sigma_{\hat{\ell}}\,F_{\hat{j}}\,\frac{\partial
    D^k_{\hat{k}}}{\partial t}\,\frac{\partial}{\partial
    x^k}\nonumber\\&&+ \frac{1}{4 m
  }\,\gamma^{\hat{0}}\,i\,\epsilon^{\hat{j}\hat{k}\hat{\ell}}\,
  \Sigma_{\hat{\ell}}\, D^k_{\hat{k}}\,\frac{\partial}{\partial
    x^k}\Big(\frac{F_{\hat{j}}}{A}\Big)\,\frac{\partial}{\partial t} -
  \frac{1}{2 m
    A}\,\gamma^{\hat{0}}\,\eta^{\hat{j}\hat{k}}\,G_{\hat{j}}
  D^k_{\hat{k}}\, \frac{\partial}{\partial x^k} - \frac{1}{4 m
  }\,\gamma^{\hat{0}}\,\eta^{\hat{j}\hat{k}}\,
  D^k_{\hat{k}}\,\frac{\partial}{\partial
    x^k}\Big(\frac{G_{\hat{j}}}{A}\Big)\nonumber\\
&& + \frac{1}{4 m
  }\,\gamma^{\hat{0}}\,i\,\epsilon^{\hat{j}\hat{k}\hat{\ell}}\,
  \Sigma_{\hat{\ell}}\, D^k_{\hat{k}}\,\frac{\partial}{\partial
    x^k}\Big(\frac{G_{\hat{j}}}{A}\Big) - \frac{1}{2 m
    A}\,i\gamma^{\hat{0}}\,\Sigma^{\hat{k}}\,K\,D^k_{\hat{k}}\,
  \frac{\partial}{\partial x^k} - \frac{1}{4 m
  }\,i\,\gamma^{\hat{0}}\,\Sigma^{\hat{k}}\,D^k_{\hat{k}}\,\frac{\partial}{\partial
    x^k}\Big(\frac{K}{A}\Big)\nonumber\\
\end{eqnarray*}
\begin{eqnarray}\label{eq:A.7}
&& - \frac{1}{2 m
    A}\,\gamma^{\hat{0}}\,\eta^{\hat{j}\hat{k}}\,
  D^j_{\hat{j}}\,F_{\hat{k}} \frac{\partial^2 }{\partial t\partial
    x^j} - \frac{1}{4 m
    A}\,\gamma^{\hat{0}}\,\eta^{\hat{j}\hat{k}}\,D^j_{\hat{j}}\,\frac{\partial
    F_{\hat{k}}}{\partial x^j}\,\frac{\partial}{\partial t} -
  \frac{1}{4 m
  }\,\gamma^{\hat{0}}\,\eta^{\hat{j}\hat{k}}\,F_{\hat{k}}\,\frac{\partial
  }{\partial
    t}\Big(\frac{D^j_{\hat{j}}}{A}\Big)\,\frac{\partial}{\partial x^j}
  - \frac{1}{4 m
    A}\,\gamma^{\hat{0}}\,i\,\epsilon^{\hat{j}\hat{k}\hat{\ell}}\,
  \Sigma_{\hat{\ell}}\,D^j_{\hat{j}}\,\frac{\partial
    F_{\hat{k}}}{\partial x^j}\,\frac{\partial}{\partial
    t}\nonumber\\
&& + \frac{1}{4 m
  }\,\gamma^{\hat{0}}\,i\,\epsilon^{\hat{j}\hat{k}\hat{\ell}}\,
  \Sigma_{\hat{\ell}}\,F_{\hat{k}}\,\frac{\partial }{\partial
    t}\Big(\frac{D^j_{\hat{j}}}{A}\Big)\,\frac{\partial}{\partial x^j}
  - \frac{1}{2 m A}\,\gamma^{\hat{0}}\,\eta^{\hat{j}\hat{k}}\,
  F_{\hat{j}}\,F_{\hat{k}} \frac{\partial^2 }{\partial t^2} -
  \frac{1}{4 m
    A}\,\gamma^{\hat{0}}\,\eta^{\hat{j}\hat{k}}\,F_{\hat{j}}\,\frac{\partial
    F_{\hat{k}}}{\partial t}\,\frac{\partial}{\partial t}\nonumber\\
&&
  - \frac{1}{4 m
  }\,\gamma^{\hat{0}}\,\eta^{\hat{j}\hat{k}}\,F_{\hat{k}}\,\frac{\partial
  }{\partial
    t}\Big(\frac{F_{\hat{j}}}{A}\Big)\,\frac{\partial}{\partial t} -
  \frac{1}{4 m
    A}\,\gamma^{\hat{0}}\,i\,\epsilon^{\hat{j}\hat{k}\hat{\ell}}\,
  \Sigma_{\hat{\ell}}\,F_{\hat{j}}\,\frac{\partial
    F_{\hat{k}}}{\partial t}\,\frac{\partial}{\partial t} + \frac{1}{4
    m }\,\gamma^{\hat{0}}\,i\,\epsilon^{\hat{j}\hat{k}\hat{\ell}}\,
  \Sigma_{\hat{\ell}}\,F_{\hat{k}}\,\frac{\partial}{\partial
    t}\Big(\frac{F_{\hat{j}}}{A}\Big)\, \frac{\partial}{\partial t} -
  \frac{1}{2 m
    A}\,\gamma^{\hat{0}}\,\eta^{\hat{j}\hat{k}}\,G_{\hat{j}}\,F_{\hat{k}}\,
  \frac{\partial}{\partial t}\nonumber\\&& -
  \frac{1}{4m}\,\gamma^{\hat{0}}\,\eta^{\hat{j}\hat{k}}\,F_{\hat{k}}\,
  \frac{\partial}{\partial t}\Big(\frac{G_{\hat{j}}}{A}\Big) +
  \frac{1}{4m}\,\gamma^{\hat{0}}\,i\,\epsilon^{\hat{j}\hat{k}\hat{\ell}}\,
  \Sigma_{\hat{\ell}}\,F_{\hat{k}}\, \frac{\partial}{\partial
    t}\Big(\frac{G_{\hat{j}}}{A}\Big) - \frac{1}{2 m
    A}\,i\,\gamma^{\hat{0}}\,\Sigma^{\hat{k}}\,K\,F_{\hat{k}}\,
  \frac{\partial}{\partial t} - \frac{1}{4 m
  }\,i\,\gamma^{\hat{0}}\,\Sigma^{\hat{k}}\,F_{\hat{k}}\,
  \frac{\partial}{\partial t}\Big(\frac{K}{A}\Big) \nonumber\\
&& -
  \frac{1}{2 m
    A}\,\gamma^{\hat{0}}\,\eta^{\hat{j}\hat{k}}\,D^j_{\hat{j}}\,G_{\hat{k}}\,
  \frac{\partial}{\partial x^j} - \frac{1}{4 m
    A}\,\gamma^{\hat{0}}\,\eta^{\hat{j}\hat{k}}\,D^j_{\hat{j}}\,
  \frac{\partial G_{\hat{k}}}{\partial x^j} - \frac{1}{4 m
    A}\,\gamma^{\hat{0}}\,i\,\epsilon^{\hat{j}\hat{k}\hat{\ell}}\,
  \Sigma_{\hat{\ell}}\,D^j_{\hat{j}}\, \frac{\partial
    G_{\hat{k}}}{\partial x^j} - \frac{1}{2 m
    A}\,\gamma^{\hat{0}}\,\eta^{\hat{j}\hat{k}}\,F_{\hat{j}}\,G_{\hat{k}}\,
  \frac{\partial}{\partial t}\nonumber\\
&& - \frac{1}{4 m
    A}\,\gamma^{\hat{0}}\,\eta^{\hat{j}\hat{k}}\,F_{\hat{j}}\,
  \frac{\partial G_{\hat{k}}}{\partial t} - \frac{1}{4 m
    A}\,\gamma^{\hat{0}}\,i\,\epsilon^{\hat{j}\hat{k}\hat{\ell}}\,
  \Sigma_{\hat{\ell}}\,F_{\hat{j}}\, \frac{\partial
    G_{\hat{k}}}{\partial t} - \frac{1}{2 m
    A}\,\gamma^{\hat{0}}\,\eta^{\hat{j}\hat{k}}\,G_{\hat{j}}\,G_{\hat{k}}
  - \frac{1}{2 m
    A}\,i\,\gamma^{\hat{0}}\,\Sigma^{\hat{k}}\,K\,G_{\hat{k}}
  \nonumber\\&& - \frac{1}{2 m
    A}\,i\,\gamma^{\hat{0}}\,\Sigma^{\hat{j}}\,D^j_{\hat{j}}\,K\,
  \frac{\partial}{\partial x^j} - \frac{1}{4 m
    A}\,i\,\gamma^{\hat{0}}\,\Sigma^{\hat{j}}\,D^j_{\hat{j}}\,
  \frac{\partial K}{\partial x^j} - \frac{1}{2 m
    A}\,i\,\gamma^{\hat{0}}\,\Sigma^{\hat{j}}\,F_{\hat{j}}\,K\,
  \frac{\partial}{\partial t} - \frac{1}{4 m
    A}\,i\,\gamma^{\hat{0}}\,\Sigma^{\hat{j}}\,F_{\hat{j}}\,
  \frac{\partial K}{\partial t}\nonumber\\
&& - \frac{1}{2 m
    A}\,i\,\gamma^{\hat{0}}\,\Sigma^{\hat{j}}\,G_{\hat{j}}\,K -
  \frac{1}{2 m A}\,\gamma^{\hat{0}}\,K^2 - \frac{1}{4 m
    A^2}\,\gamma^{\hat{0}}\,\eta^{\hat{j}\hat{k}}\,D^j_{\hat{j}}\,D^k_{\hat{k}}\,\frac{\partial
    A}{\partial x^k}\, \frac{\partial}{\partial x^j} - \frac{1}{8 m
    A}\,\gamma^{\hat{0}}\,\eta^{\hat{j}\hat{k}}\,D^j_{\hat{j}}\,
  \frac{\partial}{\partial
    x^j}\Big(\frac{D^k_{\hat{k}}}{A}\,\frac{\partial A}{\partial
    x^k}\Big)\nonumber\\&& - \frac{1}{8 m
    A}\,\gamma^{\hat{0}}\,i\,\epsilon^{\hat{j}\hat{k}\hat{\ell}}\,
  \Sigma_{\hat{\ell}}\,D^j_{\hat{j}}\, \frac{\partial}{\partial
    x^j}\Big(\frac{D^k_{\hat{k}}}{A}\,\frac{\partial A}{\partial
    x^k}\Big) - \frac{1}{4 m
    A^2}\,\gamma^{\hat{0}}\,\eta^{\hat{j}\hat{k}}\,F_{\hat{j}}\,D^k_{\hat{k}}\,
  \frac{\partial A}{\partial x^k}\, \frac{\partial}{\partial t} -
  \frac{1}{8 m
    A}\,\gamma^{\hat{0}}\,\eta^{\hat{j}\hat{k}}\,F_{\hat{j}}\,
  \frac{\partial}{\partial
    t}\Big(\frac{D^k_{\hat{k}}}{A}\,\frac{\partial A}{\partial
    x^k}\Big)\nonumber\\
&& - \frac{1}{8 m
    A}\,\gamma^{\hat{0}}\,i\,\epsilon^{\hat{j}\hat{k}\hat{\ell}}\,
  \Sigma_{\hat{\ell}}\,F_{\hat{j}}\, \frac{\partial}{\partial
    t}\Big(\frac{D^k_{\hat{k}}}{A}\,\frac{\partial A}{\partial
    x^k}\Big) - \frac{1}{4 m
    A^2}\,\gamma^{\hat{0}}\,\eta^{\hat{j}\hat{k}}\,
  G_{\hat{j}}\,D^k_{\hat{k}}\,\frac{\partial A}{\partial x^k} -
  \frac{1}{4 m A^2}\,i\,\gamma^{\hat{0}}\,\Sigma^{\hat{k}}\,
  K\,D^k_{\hat{k}}\,\frac{\partial A}{\partial x^k}\nonumber\\
&& -
  \frac{1}{4 m
    A^2}\,\gamma^{\hat{0}}\,\eta^{\hat{j}\hat{k}}\,D^j_{\hat{j}}\,F_{\hat{k}}\,
  \frac{\partial A}{\partial t}\, \frac{\partial}{\partial x^j} -
  \frac{1}{8 m
    A}\,\gamma^{\hat{0}}\,\eta^{\hat{j}\hat{k}}\,D^j_{\hat{j}}\,
  \frac{\partial}{\partial
    x^j}\Big(\frac{F_{\hat{k}}}{A}\,\frac{\partial A}{\partial t}\Big)
  - \frac{1}{8 m
    A}\,\gamma^{\hat{0}}\,i\,\epsilon^{\hat{j}\hat{k}\hat{\ell}}\,
  \Sigma_{\hat{\ell}}\,D^j_{\hat{j}}\, \frac{\partial}{\partial
    x^j}\Big(\frac{F_{\hat{k}}}{A}\,\frac{\partial A}{\partial
    t}\Big)\nonumber\\
&& - \frac{1}{4 m
    A^2}\,\gamma^{\hat{0}}\,\eta^{\hat{j}\hat{k}}\,F_{\hat{j}}\,F_{\hat{k}}\,
  \frac{\partial A}{\partial t}\, \frac{\partial}{\partial
    t} - \frac{1}{8 m
    A}\,\gamma^{\hat{0}}\,\eta^{\hat{j}\hat{k}}\,F_{\hat{j}}\,
  \frac{\partial}{\partial
    t}\Big(\frac{F_{\hat{k}}}{A}\,\frac{\partial A}{\partial t}\Big) -
  \frac{1}{8 m
    A}\,\gamma^{\hat{0}}\,i\,\epsilon^{\hat{j}\hat{k}\hat{\ell}}\,
  \Sigma_{\hat{\ell}}\,F_{\hat{j}}\, \frac{\partial}{\partial
    t}\Big(\frac{F_{\hat{k}}}{A}\,\frac{\partial A}{\partial t}\Big)
  \nonumber\\
&&- \frac{1}{4 m
    A^2}\,\gamma^{\hat{0}}\,\eta^{\hat{j}\hat{k}}\,G_{\hat{j}}\,F_{\hat{k}}\,
  \frac{\partial A}{\partial t} - \frac{1}{4 m
    A^2}\,i\,\gamma^{\hat{0}}\,\Sigma^{\hat{k}}\,K\,F_{\hat{k}}\,\frac{\partial
    A}{\partial t}.
\end{eqnarray}
The Hamilton operator ${\rm H}_1$ we decompose into two parts ${\rm
  H}_1 = {\rm H}_{1\it even} + {\rm H}_{1\it odd}$, where the
operators ${\rm H}_{1\it even}$ and ${\rm H}_{1\it odd}$ are equal to
\begin{eqnarray*}
&&{\rm H}_{1\it even} = A\,\gamma^{\hat{0}} m + B + C^{\hat{\ell}}
  \Sigma_{\hat{\ell}} + L^j\,i\,\frac{\partial}{\partial x^j} +
  \frac{1}{2 m
    A}\,\gamma^{\hat{0}}\,\eta^{\hat{j}\hat{k}}\,D^j_{\hat{j}}D^k_{\hat{k}}\,
  \frac{\partial^2}{\partial x^j \partial x^k}\nonumber\\ &&+
  \frac{1}{4 m
    A}\,\gamma^{\hat{0}}\,\eta^{\hat{j}\hat{k}}\,D^j_{\hat{j}}\frac{\partial
    D^k_{\hat{k}}}{\partial x^j}\,\frac{\partial}{\partial x^k} +
  \frac{1}{4
    m}\,\gamma^{\hat{0}}\,\eta^{\hat{j}\hat{k}}\,D^k_{\hat{k}}\,
  \frac{\partial }{\partial
    x^k}\Big(\frac{D^j_{\hat{j}}}{A}\Big)\,\frac{\partial}{\partial
    x^j} + \frac{1}{4 m
    A}\,\gamma^{\hat{0}}\,i\,\epsilon^{\hat{j}\hat{k}\hat{\ell}}\,
  \Sigma_{\hat{\ell}}\,D^j_{\hat{j}}\frac{\partial
    D^k_{\hat{k}}}{\partial x^j}\,\frac{\partial}{\partial x^k}
  \nonumber\\ 
&&- \frac{1}{4
    m}\,\gamma^{\hat{0}}\,i\,\epsilon^{\hat{j}\hat{k}\hat{\ell}}\,
  \Sigma_{\hat{\ell}}\,D^k_{\hat{k}}\, \frac{\partial }{\partial
    x^k}\Big(\frac{D^j_{\hat{j}}}{A}\Big)\,\frac{\partial}{\partial
    x^j} + \frac{1}{2 m
    A}\,\gamma^{\hat{0}}\,\eta^{\hat{j}\hat{k}}\,F_{\hat{j}}
  D^k_{\hat{k}}\,\frac{\partial^2}{\partial x^k \partial t} +
  \frac{1}{4 m
    A}\,\gamma^{\hat{0}}\,\eta^{\hat{j}\hat{k}}\,F_{\hat{j}}\frac{\partial
    D^k_{\hat{k}}}{\partial t}\,\frac{\partial}{\partial
    x^k}\nonumber\\ 
&& + \frac{1}{4
    m}\,\gamma^{\hat{0}}\,\eta^{\hat{j}\hat{k}}\,D^k_{\hat{k}}
  \frac{\partial}{\partial x^k}
  \Big(\frac{F_{\hat{j}}}{A}\Big)\,\frac{\partial }{\partial t} +
  \frac{1}{4 m
    A}\,\gamma^{\hat{0}}\,i\,\epsilon^{\hat{j}\hat{k}\hat{\ell}}\,\Sigma_{\hat{\ell}}\,F_{\hat{j}}\,\frac{\partial
    D^k_{\hat{k}}}{\partial t}\,\frac{\partial}{\partial
    x^k}\nonumber\\ 
&& - \frac{1}{4
    m}\,\gamma^{\hat{0}}\,i\,\epsilon^{\hat{j}\hat{k}\hat{\ell}}\,
  \Sigma_{\hat{\ell}}\,D^k_{\hat{k}}\,\frac{\partial}{\partial
    x^k}\Big(\frac{F_{\hat{j}}}{A}\Big)\,\frac{\partial }{\partial t}
  + \frac{1}{2 m
    A}\,\gamma^{\hat{0}}\,\eta^{\hat{j}\hat{k}}\,G_{\hat{j}}\,D^k_{\hat{k}}\,
  \frac{\partial}{\partial x^k} + \frac{1}{4 m
  }\,\gamma^{\hat{0}}\,\eta^{\hat{j}\hat{k}}\,D^k_{\hat{k}}\,
  \frac{\partial}{\partial
    x^k}\Big(\frac{G_{\hat{j}}}{A}\Big)\nonumber\\ 
&& - \frac{1}{4 m
  }\,\gamma^{\hat{0}}\,i\,\epsilon^{\hat{j}\hat{k}\hat{\ell}}\,
  \Sigma_{\hat{\ell}}\,D^k_{\hat{k}}\,\frac{\partial}{\partial
    x^k}\Big(\frac{G_{\hat{j}}}{A}\Big) + \frac{1}{2 m
    A}\,i\,\gamma^{\hat{0}}\Sigma^{\hat{k}}\,K\,
  D^k_{\hat{k}}\,\frac{\partial }{\partial x^k} + \frac{1}{4
    m}\,i\,\gamma^{\hat{0}}\Sigma^{\hat{k}}\,D^k_{\hat{k}}\,
  \frac{\partial}{\partial x^k}\Big(\frac{K}{A}\Big)\nonumber\\ && +
  \frac{1}{2 m
    A}\,\gamma^{\hat{0}}\,\eta^{\hat{j}\hat{k}}\,D^j_{\hat{j}}\,F_{\hat{k}}\,
  \frac{\partial^2}{\partial x^j\partial t} + \frac{1}{4 m
    A}\,\gamma^{\hat{0}}\,\eta^{\hat{j}\hat{k}}\,D^j_{\hat{j}}\,\frac{\partial
    F_{\hat{k}}}{\partial x^j}\,\frac{\partial}{\partial t} +
  \frac{1}{4
    m}\,\gamma^{\hat{0}}\,\eta^{\hat{j}\hat{k}}\,F_{\hat{k}}\,
  \frac{\partial}{\partial
    t}\Big(\frac{D^j_{\hat{j}}}{A}\Big)\,\frac{\partial }{\partial
    x^j}\nonumber\\ 
&& + \frac{1}{4 m
    A}\,\gamma^{\hat{0}}\,i\,\epsilon^{\hat{j}\hat{k}\hat{\ell}}\,
  \Sigma_{\hat{\ell}}\,D^j_{\hat{j}}\,\frac{\partial
    F_{\hat{k}}}{\partial x^j}\,\frac{\partial}{\partial t} -
  \frac{1}{4
    m}\,\gamma^{\hat{0}}\,i\,\epsilon^{\hat{j}\hat{k}\hat{\ell}}\,
  \Sigma_{\hat{\ell}}\,F_{\hat{k}}\,\frac{\partial}{\partial
    t}\Big(\frac{D^j_{\hat{j}}}{A}\Big)\,\frac{\partial }{\partial
    x^j}+ \frac{1}{2 m
    A}\,\gamma^{\hat{0}}\,\eta^{\hat{j}\hat{k}}\,F_{\hat{j}}\,F_{\hat{k}}\,
  \frac{\partial^2}{\partial t^2}\nonumber\\ && + \frac{1}{4 m
    A}\,\gamma^{\hat{0}}\,\eta^{\hat{j}\hat{k}}\,F_{\hat{j}}\,\frac{\partial
    F_{\hat{k}}}{\partial t}\,\frac{\partial}{\partial t} + \frac{1}{4
    m
  }\,\gamma^{\hat{0}}\,\eta^{\hat{j}\hat{k}}\,F_{\hat{k}}\,\frac{\partial
  }{\partial
    t}\Big(\frac{F_{\hat{j}}}{A}\Big)\,\frac{\partial}{\partial t} +
  \frac{1}{4 m
    A}\,\gamma^{\hat{0}}\,i\,\epsilon^{\hat{j}\hat{k}\hat{\ell}}\,
  \Sigma_{\hat{\ell}}\,F_{\hat{j}}\,\frac{\partial
    F_{\hat{k}}}{\partial t}\,\frac{\partial}{\partial
    t}\nonumber\\ 
&& - \frac{1}{4 m
  }\,\gamma^{\hat{0}}\,i\,\epsilon^{\hat{j}\hat{k}\hat{\ell}}\,
  \Sigma_{\hat{\ell}}\,F_{\hat{k}}\,\frac{\partial }{\partial
    t}\Big(\frac{F_{\hat{j}}}{A}\Big)\,\frac{\partial}{\partial t} +
  \frac{1}{2 m A}\,\gamma^{\hat{0}}\,\eta^{\hat{j}\hat{k}}\,
  G_{\hat{j}}\,F_{\hat{k}}\, \frac{\partial}{\partial t} + \frac{1}{4
    m
  }\,\gamma^{\hat{0}}\,\eta^{\hat{j}\hat{k}}\,F_{\hat{k}}\,\frac{\partial
  }{\partial t}\Big(\frac{G_{\hat{j}}}{A}\Big)\nonumber\\ 
 \end{eqnarray*}
\begin{eqnarray}\label{eq:A.8}
&& -
  \frac{1}{4 m }\,\gamma^{\hat{0}}\,
  i\,\epsilon^{\hat{j}\hat{k}\hat{\ell}}\,
  \Sigma_{\hat{\ell}}\,F_{\hat{k}}\,\frac{\partial }{\partial
    t}\Big(\frac{G_{\hat{j}}}{A}\Big) + \frac{1}{2 m
    A}\,i\,\gamma^{\hat{0}}\,\Sigma^{\hat{k}}\,K\, F_{\hat{k}}\,
  \frac{\partial}{\partial t} + \frac{1}{4 m
  }\,i\,\gamma^{\hat{0}}\,\Sigma^{\hat{k}}\,F_{\hat{k}}\,\frac{\partial}{\partial
    t}\Big(\frac{K}{A}\Big)\nonumber\\
&& + \frac{1}{2 m
    A}\,\gamma^{\hat{0}}\,\eta^{\hat{j}\hat{k}}\,G_{\hat{k}}\,D^j_{\hat{j}}
  \frac{\partial}{\partial x^j} + \frac{1}{4 m
    A}\,\gamma^{\hat{0}}\,\eta^{\hat{j}\hat{k}}\,D^j_{\hat{j}}\frac{\partial
    G_{\hat{k}}}{\partial x^j} + \frac{1}{4 m
    A}\,\gamma^{\hat{0}}\,i\,\epsilon^{\hat{j}\hat{k}\hat{\ell}}\,
  \Sigma_{\hat{\ell}}\,D^j_{\hat{j}}\frac{\partial
    G_{\hat{k}}}{\partial x^j}\nonumber\\&& + \frac{1}{2 m
    A}\,\gamma^{\hat{0}}\,\eta^{\hat{j}\hat{k}}\,F_{\hat{j}}\,G_{\hat{k}}\,
  \frac{\partial }{\partial t} + \frac{1}{4 m
    A}\,\gamma^{\hat{0}}\,\eta^{\hat{j}\hat{k}}\,F_{\hat{j}}\,\frac{\partial
    G_{\hat{k}} }{\partial t} + \frac{1}{4 m
    A}\,\gamma^{\hat{0}}\,i\,\epsilon^{\hat{j}\hat{k}\hat{\ell}}\,
  \Sigma_{\hat{\ell}}\,F_{\hat{j}}\,\frac{\partial G_{\hat{k}}
  }{\partial t}\nonumber\\
&& + \frac{1}{2 m
    A}\,\gamma^{\hat{0}}\,\eta^{\hat{j}\hat{k}}\,G_{\hat{j}}\,G_{\hat{k}}
  + \frac{1}{2 m A}\,i\,\gamma^{\hat{0}}\,\Sigma^{\hat{k}}
  K\,G_{\hat{k}} + \frac{1}{2 m
    A}\,i\,\gamma^{\hat{0}}\,\Sigma^{\hat{j}}\,K\,D^j_{\hat{j}}\,
  \frac{\partial}{\partial x^j} + \frac{1}{4 m
    A}\,i\,\gamma^{\hat{0}}\,\Sigma^{\hat{j}}\,D^j_{\hat{j}}\,
  \frac{\partial K}{\partial x^j}\nonumber\\
&& + \frac{1}{2 m
    A}\,i\,\gamma^{\hat{0}}\,\Sigma^{\hat{j}}\,K\,F_{\hat{j}}\,
  \frac{\partial}{\partial t} + \frac{1}{4 m
    A}\,i\,\gamma^{\hat{0}}\,\Sigma^{\hat{j}}\,F_{\hat{j}}\,
  \frac{\partial K}{\partial t}+ \frac{1}{2 m
    A}\,i\,\gamma^{\hat{0}}\,\Sigma^{\hat{j}}\,K\,G_{\hat{j}} +
  \frac{1}{2 m A}\,\gamma^{\hat{0}}\,K^2 \nonumber\\&&-\frac{1}{4 m
    A^2}\,\gamma^{\hat{0}}\,\eta^{\hat{j}\hat{k}}\,D^j_{\hat{j}}\,D^k_{\hat{k}}\,\frac{\partial
    A}{\partial x^k}\, \frac{\partial}{\partial x^j} - \frac{1}{8 m
    A}\,\gamma^{\hat{0}}\,\eta^{\hat{j}\hat{k}}\,D^j_{\hat{j}}\,
  \frac{\partial}{\partial
    x^j}\Big(\frac{D^k_{\hat{k}}}{A}\,\frac{\partial A}{\partial
    x^k}\Big)\nonumber\\
&& - \frac{1}{8 m
    A}\,\gamma^{\hat{0}}\,i\,\epsilon^{\hat{j}\hat{k}\hat{\ell}}\,
  \Sigma_{\hat{\ell}}\,D^j_{\hat{j}}\, \frac{\partial}{\partial
    x^j}\Big(\frac{D^k_{\hat{k}}}{A}\,\frac{\partial A}{\partial
    x^k}\Big) - \frac{1}{4 m
    A^2}\,\gamma^{\hat{0}}\,\eta^{\hat{j}\hat{k}}\,F_{\hat{j}}\,D^k_{\hat{k}}\,
\frac{\partial
    A}{\partial x^k}\, \frac{\partial}{\partial t}\nonumber\\
&& -
  \frac{1}{8 m
    A}\,\gamma^{\hat{0}}\,\eta^{\hat{j}\hat{k}}\,F_{\hat{j}}\,
  \frac{\partial}{\partial
    t}\Big(\frac{D^k_{\hat{k}}}{A}\,\frac{\partial A}{\partial
    x^k}\Big) - \frac{1}{8 m
    A}\,\gamma^{\hat{0}}\,i\,\epsilon^{\hat{j}\hat{k}\hat{\ell}}\,
  \Sigma_{\hat{\ell}}\,F_{\hat{j}}\, \frac{\partial}{\partial
    t}\Big(\frac{D^k_{\hat{k}}}{A}\,\frac{\partial A}{\partial
    x^k}\Big)\nonumber\\
&& - \frac{1}{4 m
    A^2}\,\gamma^{\hat{0}}\,\eta^{\hat{j}\hat{k}}\,
  G_{\hat{j}}\,D^k_{\hat{k}}\,\frac{\partial A}{\partial x^k} -
  \frac{1}{4 m A^2}\,i\,\gamma^{\hat{0}}\,\Sigma^{\hat{k}}\,
  K\,D^k_{\hat{k}}\,\frac{\partial A}{\partial x^k}\nonumber\\
&& -
  \frac{1}{4 m
    A^2}\,\gamma^{\hat{0}}\,\eta^{\hat{j}\hat{k}}\,D^j_{\hat{j}}\,F_{\hat{k}}\,
  \frac{\partial A}{\partial t}\, \frac{\partial}{\partial x^j} -
  \frac{1}{8 m
    A}\,\gamma^{\hat{0}}\,\eta^{\hat{j}\hat{k}}\,D^j_{\hat{j}}\,
  \frac{\partial}{\partial
    x^j}\Big(\frac{F_{\hat{k}}}{A}\,\frac{\partial A}{\partial t}\Big)
  \nonumber\\
&& - \frac{1}{8 m
    A}\,\gamma^{\hat{0}}\,i\,\epsilon^{\hat{j}\hat{k}\hat{\ell}}\,
  \Sigma_{\hat{\ell}}\,D^j_{\hat{j}}\, \frac{\partial}{\partial
    x^j}\Big(\frac{F_{\hat{k}}}{A}\,\frac{\partial A}{\partial t}\Big)
  - \frac{1}{4 m
    A^2}\,\gamma^{\hat{0}}\,\eta^{\hat{j}\hat{k}}\,F_{\hat{j}}\,F_{\hat{k}}\,
  \frac{\partial A}{\partial t}\, \frac{\partial}{\partial
    t}\nonumber\\ 
&&- \frac{1}{8 m
    A}\,\gamma^{\hat{0}}\,\eta^{\hat{j}\hat{k}}\,F_{\hat{j}}\,
  \frac{\partial}{\partial
    t}\Big(\frac{F_{\hat{k}}}{A}\,\frac{\partial A}{\partial t}\Big) -
  \frac{1}{8 m
    A}\,\gamma^{\hat{0}}\,i\,\epsilon^{\hat{j}\hat{k}\hat{\ell}}\,
  \Sigma_{\hat{\ell}}\,F_{\hat{j}}\, \frac{\partial}{\partial
    t}\Big(\frac{F_{\hat{k}}}{A}\,\frac{\partial A}{\partial t}\Big)
  \nonumber\\&&- \frac{1}{4 m
    A^2}\,\gamma^{\hat{0}}\,\eta^{\hat{j}\hat{k}}\,G_{\hat{j}}\,F_{\hat{k}}\,
  \frac{\partial A}{\partial t} - \frac{1}{4 m
    A^2}\,i\,\gamma^{\hat{0}}\,\Sigma^{\hat{k}}\,K\,F_{\hat{k}}\,\frac{\partial
    A}{\partial t}
\end{eqnarray}
and
\begin{eqnarray}\label{eq:A.9}
{\rm H}_{1\it odd} &=& -
\frac{1}{2}\,i\,\gamma^{\hat{0}}\gamma^{\hat{j}}\frac{D^j_{\hat{j}}}{A}\,
\frac{\partial A}{\partial x^j} -
\frac{1}{2}\,i\,\gamma^{\hat{0}}\gamma^{\hat{j}}\frac{F_{\hat{j}}}{A}\,
\frac{\partial A}{\partial t}\nonumber\\ &&+
\frac{1}{2m A}\,i\,\gamma^{\hat{j}}\,
D^j_{\hat{j}}\,\frac{\partial B}{\partial x^j} +
\frac{1}{2m A}\,i\,\gamma^{\hat{j}}\,
F_{\hat{j}}\,\frac{\partial B}{\partial t} - \frac{1}{2m
  A}\,\eta^{\hat{j}\hat{\ell}}D^j_{\hat{j}}\frac{\partial
  C_{\hat{\ell}}}{\partial x^j}\,i\,\gamma^{\hat{0}}\gamma^5 +
\frac{1}{2m
  A}\,\epsilon^{\hat{j}\hat{\ell}\hat{k}}\,\gamma_{\hat{k}}\,D^j_{\hat{j}}\,
\frac{\partial C_{\hat{\ell}}}{\partial x^j}\nonumber\\ 
&&+ \frac{1}{m
  A}\,\epsilon^{\hat{j}\hat{\ell}\hat{k}}\,\gamma_{\hat{k}}\,D^j_{\hat{j}}
C_{\hat{\ell}}\,\frac{\partial }{\partial x^j} - \frac{1}{2m
  A}\,\eta^{\hat{j}\hat{\ell}}F_{\hat{j}}\frac{\partial
  C_{\hat{\ell}}}{\partial t}\,i\,\gamma^{\hat{0}}\gamma^5 +
\frac{1}{2m
  A}\,\epsilon^{\hat{j}\hat{\ell}\hat{k}}\,\gamma_{\hat{k}}\,F_{\hat{j}}\,
\frac{\partial C_{\hat{\ell}}}{\partial t}\nonumber\\ &&+ \frac{1}{m
  A}\,\epsilon^{\hat{j}\hat{\ell}\hat{k}}\,\gamma_{\hat{k}}\,F_{\hat{j}}
C_{\hat{\ell}}\,\frac{\partial }{\partial t} + \frac{1}{m
  A}\,\epsilon^{\hat{j}\hat{\ell}\hat{k}}\,\gamma_{\hat{k}}\,G_{\hat{j}}\,
C_{\hat{\ell}}\nonumber\\ 
&&- \frac{1}{2 m
  A}\,\gamma^{\hat{j}}\,D^j_{\hat{j}}\,\frac{\partial L^k}{\partial
  x^j}\,\frac{\partial}{\partial x^k} + \frac{1}{2
  m}\,\,\gamma^{\hat{j}}\,L^k\,\frac{\partial}{\partial
  x^k}\Big(\frac{D^j_{\hat{j}}}{A}\Big)\,\frac{\partial }{\partial
  x^j} - \frac{1}{2 m
  A}\,\gamma^{\hat{j}}\,F_{\hat{j}}\,\frac{\partial L^k}{\partial
  t}\,\frac{\partial}{\partial x^k}\nonumber\\
&& + \frac{1}{2
  m}\,\,\gamma^{\hat{j}}\,L^k\,\frac{\partial}{\partial
  x^k}\Big(\frac{F_{\hat{j}}}{A}\Big)\,\frac{\partial }{\partial t} +
\frac{1}{2 m}\,\gamma^{\hat{j}}\,L^k\,\frac{\partial}{\partial
  x^k}\Big(\frac{G_{\hat{j}}}{A}\Big) - \frac{1}{2
  m}\,L^k\,\frac{\partial}{\partial
  x^k}\Big(\frac{K}{A}\Big)\,i\,\gamma^{\hat{0}}\,\gamma^5\nonumber\\ && -
\frac{1}{2m}\,\frac{\partial}{\partial
  t}\Big(\frac{D^j_{\hat{j}}}{A}\Big)\,
\gamma^{\hat{j}}\frac{\partial}{\partial x^j} -
\frac{1}{2m}\,\frac{\partial}{\partial
  t}\Big(\frac{F_{\hat{j}}}{A}\Big)\,\gamma^{\hat{j}}\frac{\partial}{\partial
  t} - \frac{1}{2m}\,\frac{\partial}{\partial
  t}\Big(\frac{G_{\hat{j}}}{A}\Big)\,\gamma^{\hat{j}} +
\frac{1}{2m}\,\frac{\partial }{\partial
  t}\Big(\frac{K}{A}\Big)\,i\,\gamma^{\hat{0}}\gamma^5.
\end{eqnarray}
For the calculation of the Hamilton operator ${\rm H}_1 = {\rm
  H}_{1\rm even} + {\rm H}_{1\rm odd}$ we have neglected the
contributions of order $O(1/m^2)$. In order to remove the {\it odd}
operators we perform the second unitary transformation of the wave
function $\psi_1(x) = e^{\,-iS_2}\,\psi_2(x)$ and the Hamilton
operator ${\rm H}_1$:
\begin{eqnarray}\label{eq:A.10}
\hspace{-0.3in}{\rm H}_2 = e^{\,+ i S_2}\,{\rm H}_1\,e^{\,-i S_2} -
i\,e^{\,i S_2}\frac{\partial}{\partial t}e^{\,-i S_2} = {\rm H}_1 -
\frac{\partial S_2}{\partial t} + i\Big[S_2,{\rm H}_1 -
  \frac{1}{2}\,\frac{\partial S_2}{\partial t}\Big] +
\frac{i^2}{2}\,\Big[S_2,\Big[S_2,{\rm H}_1 -
    \frac{1}{3}\,\frac{\partial S_2}{\partial t}\Big]\Big] + \ldots,
\end{eqnarray}
where the operator $S_2$ is equal to $S_2 = - (i/2 m
A)\,\gamma^{\hat{0}}\,{\rm H}_{1\it odd}$.  Neglecting the
contributions of order $O(1/m^2)$ we arrive at the Hamilton operator
${\rm H}_2 = {\rm H}_{2\it even} + {\rm H}_{2\it odd}$, where
\begin{eqnarray*}
{\rm H}_{2\it even} &=&{\rm H}_{1\it even} +
\frac{1}{4mA^3}\,\gamma^{\hat{0}}\,\eta^{\hat{j}\hat{k}}\,D^j_{\hat{j}}
D^k_{\hat{k}}\,\frac{\partial A}{\partial x^j}\, \frac{\partial
  A}{\partial x^k} +
\frac{1}{4mA^3}\,\gamma^{\hat{0}}\,\eta^{\hat{j}\hat{k}}\,D^j_{\hat{j}}
F_{\hat{k}}\,\frac{\partial A}{\partial x^j}\, \frac{\partial
  A}{\partial t }\nonumber\\
\end{eqnarray*}
\begin{eqnarray}\label{eq:A.11}
 &+&
\frac{1}{4mA^3}\,\gamma^{\hat{0}}\,\eta^{\hat{j}\hat{k}}\,
F_{\hat{j}}\,D^k_{\hat{k}}\,\frac{\partial A}{\partial t}\,
\frac{\partial A}{\partial x^k }+
\frac{1}{4mA^3}\,\gamma^{\hat{0}}\,\eta^{\hat{j}\hat{k}}\,F_{\hat{j}}
F_{\hat{k}}\,\frac{\partial A}{\partial t}\, \frac{\partial
  A}{\partial t},\nonumber\\ 
{\rm H}_{2\it odd} &=& - \frac{1}{2 m
  A^2}\,\epsilon^{\hat{j}\hat{\ell}\hat{k}}\,\gamma_{\hat{k}}\,D^j_{\hat{j}}
C_{\hat{\ell}}\,\frac{\partial A}{\partial x^j} - \frac{1}{2 m
  A^2}\,\epsilon^{\hat{j}\hat{\ell}\hat{k}}\,\gamma_{\hat{k}}\,F_{\hat{j}}
C_{\hat{\ell}}\,\frac{\partial A}{\partial t} -
\frac{1}{4m}\,\gamma^{\hat{j}}\,L^k\,\frac{\partial}{\partial
  x^k}\Big(\frac{D^j_{\hat{j}}}{A^2}\frac{\partial A}{\partial
  x^j}\Big)\nonumber\\ &-&
\frac{1}{4m}\,\gamma^{\hat{j}}\,L^k\,\frac{\partial}{\partial
  x^k}\Big(\frac{F_{\hat{j}}}{A^2}\frac{\partial A}{\partial t}\Big) +
\frac{1}{4m}\,\gamma^{\hat{j}}\,\frac{\partial}{\partial
  t}\Big(\frac{D^j_{\hat{j}}}{A^2}\frac{\partial A}{\partial x^j}\Big)
+ \frac{1}{4m}\,\gamma^{\hat{j}}\,\frac{\partial}{\partial
  t}\Big(\frac{F_{\hat{j}}}{A^2}\frac{\partial A}{\partial t}\Big).
\end{eqnarray}
The contribution of the Hamilton operator ${\rm H}_{2\it odd}$ we
remove by the third unitary transformation of the wave function
$\psi_2(x) = e^{\,-iS_3}\,\psi_3(x)$ and the Hamilton operator ${\rm
  H}_2$:
\begin{eqnarray}\label{eq:A.12}
\hspace{-0.3in}{\rm H}_3 = e^{\,+ i S_3}\,{\rm H}_2\,e^{\,-i S_3} -
i\,e^{\,i S_3}\frac{\partial}{\partial t}e^{\,-i S_3} = {\rm H}_2 -
\frac{\partial S_3}{\partial t} + i\Big[S_3,{\rm H}_2 -
  \frac{1}{2}\,\frac{\partial S_3}{\partial t}\Big] +
\frac{i^2}{2}\,\Big[S_3,\Big[S_3,{\rm H}_2 -
    \frac{1}{3}\,\frac{\partial S_3}{\partial t}\Big]\Big] + \ldots,
\end{eqnarray}
where the operator $S_3$ is given by $S_3 = - (i/2 m
A)\,\gamma^0\,{\rm H}_{2\it odd} = O(1/m^2)$.  Neglecting the
contributions of order $O(1/m^2)$ the low--energy approximation of the
Dirac Hamilton operator for slow fermions is equal to ${\rm H}_3 =
{\rm H}_{2\it even}$, which we denote ${\rm H}_3 = {\rm H}_{2\it even}
= {\rm H}_{\rm FW}$. Thus, the Dirac equation in the low--energy
approximation takes the form
\begin{eqnarray}\label{eq:A.13}
\hspace{-0.3in}i\,\frac{\partial \psi_3(x)}{\partial t} = {\rm
  H}_{\rm FW}\,\psi_3(x),
\end{eqnarray}
where $\psi_3(x) = e^{i\,S_3}e^{i\,S_2}e^{i\,S_1}\psi'(x)$. Following
the standard procedure \cite{Foldy1950} and multiplying the both sides
of Eq.(\ref{eq:A.13}) by the matrix $(1 + \gamma^0)/2$ we arrive at
the Schr\"odinger--Pauli equation
\begin{eqnarray}\label{eq:A.14}
i\,\frac{\partial \Psi(t,\vec{r}\,)}{\partial t} =
\Big(\frac{1 + \gamma^0}{2}\Big)\,{\rm H}_{\rm FW}\,\psi_3(x) = \Big(-
\frac{1}{2m}\,\Delta + \Phi_{\rm
  eff}(t,\vec{r},\vec{\sigma}\,)\Big)\,\Psi(t,\vec{r}\,),
\end{eqnarray}
where $\Psi(t,\vec{r}\,) = {\displaystyle \frac{1 +
    \gamma^0}{2}}\,\psi_3(x)$ is the large component of the slow
  Dirac fermion wave function and $\Phi_{\rm eff}(t,
  \vec{r},\vec{\sigma}\,)$ is the effective low--energy potential
\begin{eqnarray*}
&&\Phi_{\rm eff}(t, \vec{r},\vec{\sigma}\,) = (A - 1)\, m + B +
  C^{\hat{\ell}}\sigma_{\hat{\ell}} + i\,L^j\,\frac{\partial}{\partial
    x^j} - \frac{m}{2
    A}\,\eta^{\hat{j}\hat{k}}\,F_{\hat{j}}\,F_{\hat{k}} - \frac{1}{2
    A}\,i\,\eta^{\hat{j}\hat{k}}\,F_{\hat{j}}
  D^k_{\hat{k}}\,\frac{\partial}{\partial x^k} -
  \frac{1}{4}\,i\,\eta^{\hat{j}\hat{k}}\,D^k_{\hat{k}}
  \frac{\partial}{\partial x^k}
  \Big(\frac{F_{\hat{j}}}{A}\Big)\nonumber\\ && -
  \frac{1}{4}\,\epsilon^{\hat{j}\hat{k}\hat{\ell}}\,
  \sigma_{\hat{\ell}}\,D^k_{\hat{k}}\,\frac{\partial}{\partial
    x^k}\Big(\frac{F_{\hat{j}}}{A}\Big) - \frac{1}{2
    A}\,i\,\eta^{\hat{j}\hat{k}}\,D^j_{\hat{j}}\,F_{\hat{k}}\,
  \frac{\partial}{\partial x^j} - \frac{1}{4
    A}\,i\,\eta^{\hat{j}\hat{k}}\,D^j_{\hat{j}}\,\frac{\partial
    F_{\hat{k}}}{\partial x^j} + \frac{1}{4
    A}\,\epsilon^{\hat{j}\hat{k}\hat{\ell}}\,
  \sigma_{\hat{\ell}}\,D^j_{\hat{j}}\,\frac{\partial
    F_{\hat{k}}}{\partial x^j} - \frac{1}{
    A}\,i\,\eta^{\hat{j}\hat{k}}\,F_{\hat{j}}\,F_{\hat{k}}\,
  \frac{\partial}{\partial t}\nonumber\\ && - \frac{1}{4
    A}\,i\,\eta^{\hat{j}\hat{k}}\,F_{\hat{j}}\,\frac{\partial
    F_{\hat{k}}}{\partial t} - \frac{1}{4
  }\,i\,\eta^{\hat{j}\hat{k}}\,F_{\hat{k}}\,\frac{\partial }{\partial
    t}\Big(\frac{F_{\hat{j}}}{A}\Big) + \frac{1}{4
    A}\,\epsilon^{\hat{j}\hat{k}\hat{\ell}}\,
  \sigma_{\hat{\ell}}\,F_{\hat{j}}\,\frac{\partial
    F_{\hat{k}}}{\partial t} - \frac{1}{4
  }\,\epsilon^{\hat{j}\hat{k}\hat{\ell}}\,
  \sigma_{\hat{\ell}}\,F_{\hat{k}}\,\frac{\partial }{\partial
    t}\Big(\frac{F_{\hat{j}}}{A}\Big) - \frac{1}{2
    A}\,i\,\eta^{\hat{j}\hat{k}}\, \Big(G_{\hat{j}}\,F_{\hat{k}} +
  F_{\hat{j}}\,G_{\hat{k}}\Big)\nonumber\\ && + \frac{1}{A}\,
  \sigma^{\hat{j}}\, F_{\hat{j}}\,K + \frac{1}{4
    A^2}\,i\,\eta^{\hat{j}\hat{k}}\,F_{\hat{j}}\,D^k_{\hat{k}}\,\frac{\partial
    A}{\partial x^k} + \frac{1}{4
    A^2}\,i\,\eta^{\hat{j}\hat{k}}\,F_{\hat{j}}\,F_{\hat{k}}\,
  \frac{\partial A}{\partial t} + \frac{1}{2 m
  }\,\eta^{\hat{j}\hat{k}}\,\Big(\frac{D^j_{\hat{j}}D^k_{\hat{k}}}{A}
  - \delta^j_{\hat{j}} \delta^k_{\hat{k}}\Big)\,
  \frac{\partial^2}{\partial x^j \partial x^k}\nonumber\\ && +
  \frac{1}{4 m A}\,\eta^{\hat{j}\hat{k}}\,D^j_{\hat{j}}\frac{\partial
    D^k_{\hat{k}}}{\partial x^j}\,\frac{\partial}{\partial x^k} +
  \frac{1}{4 m}\,\eta^{\hat{j}\hat{k}}\,D^k_{\hat{k}}\,\frac{\partial
  }{\partial
    x^k}\Big(\frac{D^j_{\hat{j}}}{A}\Big)\,\frac{\partial}{\partial
    x^j} + \frac{1}{4 m A}\,i\,\epsilon^{\hat{j}\hat{k}\hat{\ell}}\,
  \sigma_{\hat{\ell}}\,D^j_{\hat{j}}\frac{\partial
    D^k_{\hat{k}}}{\partial x^j}\,\frac{\partial}{\partial
    x^k}\nonumber\\ 
&& - \frac{1}{4
    m}\,i\,\epsilon^{\hat{j}\hat{k}\hat{\ell}}\,
  \sigma_{\hat{\ell}}\,D^k_{\hat{k}}\, \frac{\partial }{\partial
    x^k}\Big(\frac{D^j_{\hat{j}}}{A}\Big)\,\frac{\partial}{\partial
    x^j} + \frac{1}{2 m A}\,\eta^{\hat{j}\hat{k}}\,F_{\hat{j}}
  D^k_{\hat{k}}\,\frac{\partial^2}{\partial x^k \partial t} +
  \frac{1}{4 m A}\,\eta^{\hat{j}\hat{k}}\,F_{\hat{j}}\frac{\partial
    D^k_{\hat{k}}}{\partial t}\,\frac{\partial}{\partial x^k} +
  \frac{1}{4 m}\,\eta^{\hat{j}\hat{k}}\,D^k_{\hat{k}}
  \frac{\partial}{\partial x^k}
  \Big(\frac{F_{\hat{j}}}{A}\Big)\,\frac{\partial }{\partial
    t}\nonumber\\ 
&&+ \frac{1}{4 m
    A}\,i\,\epsilon^{\hat{j}\hat{k}\hat{\ell}}\,\sigma_{\hat{\ell}}\,F_{\hat{j}}\,\frac{\partial
    D^k_{\hat{k}}}{\partial t}\,\frac{\partial}{\partial x^k} -
  \frac{1}{4 m}\,i\,\epsilon^{\hat{j}\hat{k}\hat{\ell}}\,
  \sigma_{\hat{\ell}}\,D^k_{\hat{k}}\,\frac{\partial}{\partial
    x^k}\Big(\frac{F_{\hat{j}}}{A}\Big)\,\frac{\partial }{\partial t}
  + \frac{1}{2 m
    A}\,\eta^{\hat{j}\hat{k}}\,G_{\hat{j}}\,D^k_{\hat{k}}\,
  \frac{\partial}{\partial x^k} + \frac{1}{4 m
  }\,\eta^{\hat{j}\hat{k}}\,D^k_{\hat{k}}\, \frac{\partial}{\partial
    x^k}\Big(\frac{G_{\hat{j}}}{A}\Big)\nonumber\\ 
&& - \frac{1}{4 m
  }\,i\,\epsilon^{\hat{j}\hat{k}\hat{\ell}}\,
  \sigma_{\hat{\ell}}\,D^k_{\hat{k}}\,\frac{\partial}{\partial
    x^k}\Big(\frac{G_{\hat{j}}}{A}\Big) + \frac{1}{2 m
    A}\,i\,\sigma^{\hat{k}}\,K\, D^k_{\hat{k}}\,\frac{\partial
  }{\partial x^k} + \frac{1}{4
    m}\,i\,\sigma^{\hat{k}}\,D^k_{\hat{k}}\, \frac{\partial}{\partial
    x^k}\Big(\frac{K}{A}\Big) + \frac{1}{2 m
    A}\,\eta^{\hat{j}\hat{k}}\,D^j_{\hat{j}}\,F_{\hat{k}}\,
  \frac{\partial^2}{\partial x^j\partial t}\nonumber\\ && + \frac{1}{4
    m A}\,\eta^{\hat{j}\hat{k}}\,D^j_{\hat{j}}\,\frac{\partial
    F_{\hat{k}}}{\partial x^j}\,\frac{\partial}{\partial t} +
  \frac{1}{4
    m}\,\eta^{\hat{j}\hat{k}}\,F_{\hat{k}}\,\frac{\partial}{\partial
    t}\Big(\frac{D^j_{\hat{j}}}{A}\Big)\,\frac{\partial }{\partial
    x^j} + \frac{1}{4 m A}\,i\,\epsilon^{\hat{j}\hat{k}\hat{\ell}}\,
  \sigma_{\hat{\ell}}\,D^j_{\hat{j}}\,\frac{\partial
    F_{\hat{k}}}{\partial x^j}\,\frac{\partial}{\partial t} -
  \frac{1}{4 m}\,i\,\epsilon^{\hat{j}\hat{k}\hat{\ell}}\,
  \sigma_{\hat{\ell}}\,F_{\hat{k}}\,\frac{\partial}{\partial
    t}\Big(\frac{D^j_{\hat{j}}}{A}\Big)\,\frac{\partial }{\partial
    x^j} \nonumber\\ 
&& + \frac{1}{2 m
    A}\,\eta^{\hat{j}\hat{k}}\,F_{\hat{j}}\,F_{\hat{k}}\,
  \frac{\partial^2}{\partial t^2} + \frac{1}{4 m
    A}\,\eta^{\hat{j}\hat{k}}\,F_{\hat{j}}\,\frac{\partial
    F_{\hat{k}}}{\partial t}\,\frac{\partial}{\partial t} + \frac{1}{4
    m }\,\eta^{\hat{j}\hat{k}}\,F_{\hat{k}}\,\frac{\partial }{\partial
    t}\Big(\frac{F_{\hat{j}}}{A}\Big)\,\frac{\partial}{\partial t} +
  \frac{1}{4 m A}\,i\,\epsilon^{\hat{j}\hat{k}\hat{\ell}}\,
  \sigma_{\hat{\ell}}\,F_{\hat{j}}\,\frac{\partial
    F_{\hat{k}}}{\partial t}\,\frac{\partial}{\partial
    t}\nonumber\\ 
&& - \frac{1}{4 m
  }\,i\,\epsilon^{\hat{j}\hat{k}\hat{\ell}}\,
  \sigma_{\hat{\ell}}\,F_{\hat{k}}\,\frac{\partial }{\partial
    t}\Big(\frac{F_{\hat{j}}}{A}\Big)\,\frac{\partial}{\partial t} +
  \frac{1}{2 m A}\,\eta^{\hat{j}\hat{k}}\, G_{\hat{j}}\,F_{\hat{k}}\,
  \frac{\partial}{\partial t} + \frac{1}{4 m
  }\,\eta^{\hat{j}\hat{k}}\,F_{\hat{k}}\,\frac{\partial }{\partial
    t}\Big(\frac{G_{\hat{j}}}{A}\Big) - \frac{1}{4 m }\,
  i\,\epsilon^{\hat{j}\hat{k}\hat{\ell}}\,
  \sigma_{\hat{\ell}}\,F_{\hat{k}}\,\frac{\partial }{\partial
    t}\Big(\frac{G_{\hat{j}}}{A}\Big)\nonumber\\ 
&& + \frac{1}{2 m
    A}\,i\,\sigma^{\hat{k}}\,K\, F_{\hat{k}}\,
  \frac{\partial}{\partial t} + \frac{1}{4 m
  }\,i\,\sigma^{\hat{k}}\,F_{\hat{k}}\,\frac{\partial}{\partial
    t}\Big(\frac{K}{A}\Big) + \frac{1}{2 m
    A}\,\eta^{\hat{j}\hat{k}}\,G_{\hat{k}}\,D^j_{\hat{j}}
  \frac{\partial}{\partial x^j} + \frac{1}{4 m
    A}\,\eta^{\hat{j}\hat{k}}\,D^j_{\hat{j}}\frac{\partial
    G_{\hat{k}}}{\partial x^j} + \frac{1}{4 m
    A}\,i\,\epsilon^{\hat{j}\hat{k}\hat{\ell}}\,
  \sigma_{\hat{\ell}}\,D^j_{\hat{j}}\frac{\partial
    G_{\hat{k}}}{\partial x^j}\nonumber\\
&& + \frac{1}{2 m
    A}\,\eta^{\hat{j}\hat{k}}\,F_{\hat{j}}\,G_{\hat{k}}\,
  \frac{\partial }{\partial t} + \frac{1}{4 m
    A}\,\eta^{\hat{j}\hat{k}}\,F_{\hat{j}}\,\frac{\partial G_{\hat{k}}
  }{\partial t} + \frac{1}{4 m
    A}\,i\,\epsilon^{\hat{j}\hat{k}\hat{\ell}}\,
  \sigma_{\hat{\ell}}\,F_{\hat{j}}\,\frac{\partial G_{\hat{k}}
  }{\partial t} + \frac{1}{2 m
    A}\,\eta^{\hat{j}\hat{k}}\,G_{\hat{j}}\,G_{\hat{k}} + \frac{1}{2 m
    A}\,i\,\sigma^{\hat{k}} K\,G_{\hat{k}}\nonumber\\
\end{eqnarray*}
\begin{eqnarray}\label{eq:A.15}
&& + \frac{1}{2 m
    A}\,i\,\sigma^{\hat{j}}\,K\,D^j_{\hat{j}}\,
  \frac{\partial}{\partial x^j} + \frac{1}{4 m
    A}\,i\,\sigma^{\hat{j}}\,D^j_{\hat{j}}\, \frac{\partial
    K}{\partial x^j} + \frac{1}{2 m
    A}\,i\,\sigma^{\hat{j}}\,K\,F_{\hat{j}}\, \frac{\partial}{\partial
    t} + \frac{1}{4 m A}\,i\,\sigma^{\hat{j}}\,F_{\hat{j}}\,
  \frac{\partial K}{\partial t} + \frac{1}{2 m
    A}\,i\,\sigma^{\hat{j}}\,K\,G_{\hat{j}} + \frac{1}{2 m
    A}\,K^2\nonumber\\
&&- \frac{1}{4 m
    A^2}\,\eta^{\hat{j}\hat{k}}\,D^j_{\hat{j}}\,D^k_{\hat{k}}\,\frac{\partial
    A}{\partial x^k}\, \frac{\partial}{\partial x^j} - \frac{1}{8 m
    A}\,\eta^{\hat{j}\hat{k}}\,D^j_{\hat{j}}\,
  \frac{\partial}{\partial
    x^j}\Big(\frac{D^k_{\hat{k}}}{A}\,\frac{\partial A}{\partial
    x^k}\Big) - \frac{1}{8 m
    A}\,i\,\epsilon^{\hat{j}\hat{k}\hat{\ell}}\,
  \sigma_{\hat{\ell}}\,D^j_{\hat{j}}\, \frac{\partial}{\partial
    x^j}\Big(\frac{D^k_{\hat{k}}}{A}\,\frac{\partial A}{\partial
    x^k}\Big)\nonumber\\ && - \frac{1}{4 m
    A^2}\,\eta^{\hat{j}\hat{k}}\,F_{\hat{j}}\,D^k_{\hat{k}}\,\frac{\partial
    A}{\partial x^k}\, \frac{\partial}{\partial t} - \frac{1}{8 m
    A}\,\eta^{\hat{j}\hat{k}}\,F_{\hat{j}}\, \frac{\partial}{\partial
    t}\Big(\frac{D^k_{\hat{k}}}{A}\,\frac{\partial A}{\partial
    x^k}\Big) - \frac{1}{8 m
    A}\,i\,\epsilon^{\hat{j}\hat{k}\hat{\ell}}\,
  \sigma_{\hat{\ell}}\,F_{\hat{j}}\, \frac{\partial}{\partial
    t}\Big(\frac{D^k_{\hat{k}}}{A}\,\frac{\partial A}{\partial
    x^k}\Big)\nonumber\\ 
&& - \frac{1}{4 m
    A^2}\,\eta^{\hat{j}\hat{k}}\,
  G_{\hat{j}}\,D^k_{\hat{k}}\,\frac{\partial A}{\partial x^k} -
  \frac{1}{4 m A^2}\,i\,\sigma^{\hat{k}}\,
  K\,D^k_{\hat{k}}\,\frac{\partial A}{\partial x^k} - \frac{1}{4 m
    A^2}\,\eta^{\hat{j}\hat{k}}\,D^j_{\hat{j}}\,F_{\hat{k}}\,
  \frac{\partial A}{\partial t}\, \frac{\partial}{\partial x^j} -
  \frac{1}{8 m A}\,\eta^{\hat{j}\hat{k}}\,D^j_{\hat{j}}\,
  \frac{\partial}{\partial
    x^j}\Big(\frac{F_{\hat{k}}}{A}\,\frac{\partial A}{\partial t}\Big)
  \nonumber\\  
&& - \frac{1}{8 m
    A}\,i\,\epsilon^{\hat{j}\hat{k}\hat{\ell}}\,
  \sigma_{\hat{\ell}}\,D^j_{\hat{j}}\, \frac{\partial}{\partial
    x^j}\Big(\frac{F_{\hat{k}}}{A}\,\frac{\partial A}{\partial t}\Big)
  - \frac{1}{4 m
    A^2}\,\eta^{\hat{j}\hat{k}}\,F_{\hat{j}}\,F_{\hat{k}}\,
  \frac{\partial A}{\partial t}\, \frac{\partial}{\partial t} -
  \frac{1}{8 m A}\,\eta^{\hat{j}\hat{k}}\,F_{\hat{j}}\,
  \frac{\partial}{\partial
    t}\Big(\frac{F_{\hat{k}}}{A}\,\frac{\partial A}{\partial
    t}\Big)\nonumber\\ 
&& - \frac{1}{8 m
    A}\,i\,\epsilon^{\hat{j}\hat{k}\hat{\ell}}\,
  \sigma_{\hat{\ell}}\,F_{\hat{j}}\, \frac{\partial}{\partial
    t}\Big(\frac{F_{\hat{k}}}{A}\,\frac{\partial A}{\partial t}\Big) -
  \frac{1}{4 m A^2}\,\eta^{\hat{j}\hat{k}}\,G_{\hat{j}}\,F_{\hat{k}}\,
  \frac{\partial A}{\partial t} - \frac{1}{4 m
    A^2}\,i\,\sigma^{\hat{k}}\,K\,F_{\hat{k}}\,\frac{\partial
    A}{\partial t} +
  \frac{1}{4mA^3}\,\eta^{\hat{j}\hat{k}}\,D^j_{\hat{j}}
  D^k_{\hat{k}}\,\frac{\partial A}{\partial x^j}\, \frac{\partial
    A}{\partial x^k}\nonumber\\ 
&&+
  \frac{1}{4mA^3}\,\eta^{\hat{j}\hat{k}}\,D^j_{\hat{j}}
  F_{\hat{k}}\,\frac{\partial A}{\partial x^j}\, \frac{\partial
    A}{\partial t } + \frac{1}{4mA^3}\,\eta^{\hat{j}\hat{k}}\,
  F_{\hat{j}}\,D^k_{\hat{k}}\,\frac{\partial A}{\partial t}\,
  \frac{\partial A}{\partial x^k } +
  \frac{1}{4mA^3}\,\eta^{\hat{j}\hat{k}}\,F_{\hat{j}}
  F_{\hat{k}}\,\frac{\partial A}{\partial t}\, \frac{\partial
    A}{\partial t},
\end{eqnarray}
where $\sigma_{\hat{a}} = (- \vec{\sigma}\,)_{\hat{a}}$ and
$\sigma_{\hat{a}} = - \sigma^{\hat{a}}$.  The effective potential
Eq.(\ref{eq:A.15}) is the most general effective low--energy potential
for slow Dirac fermions in the Einstein--Cartan gravity with torsion
and chameleon, calculated to order $O(1/m)$.

We would like to note that the effective low--energy
  potential Eq.(\ref{eq:A.15}) is derived by using the
  Foldy--Wouthuysen method or the Foldy--Wouthuysen unitary
  transformations of wave functions of Dirac fermions with mass $m$,
  leading to a non--relativistic Hamilton operator, expressed in terms
  of {\it even} operators only in a from of the large fermion mass
  expansion in powers of $1/m$.  It is known that such a method of a
  reduction of the Dirac Hamilton operator to a form, containing only
  {\it even} operators, is not unique and there are some other methods
  of unitary transformations such as the Eriksen method
  \cite{Eriksen1958} and others, which were well discussed by de Vries
  \cite{Vries1970}. In section \ref{sec:conclusion} we give a short
  comparison of the Foldy--Wouthuysen with the Eriksen one only, since
  other methods of unitary transformations of a Dirac Hamilton
  operator to a form, containing only {\it even} operators, seem to be
  cumbersome when compared to the Eriksen method \cite{Vries1970}. We
  discuss also an accuracy of the Foldy--Wouthuysen representation of
  a Dirac Hamilton operator.

\end{document}